\documentclass[11pt]{article}
\usepackage{graphicx}%
\usepackage{multirow}%
\usepackage{amsmath,amssymb,amsfonts}%
\usepackage{amsthm}%
\usepackage{mathrsfs}%
\usepackage{mathtools}
\usepackage[title]{appendix}%
\usepackage{xcolor}%
\usepackage{textcomp}%
\usepackage{manyfoot}%
\usepackage{booktabs}%
\usepackage{algorithm}%
\usepackage{algorithmicx}%
\usepackage{algpseudocode}%
\usepackage{listings}%
\title{The Relativistic Gravitational Field of a Spherically Symmetric Extended Body \\
\small{Published in Physica Scripta 100, 105024 (2025).\\ Relativistic metric of extended spherically symmetric bodies, shell-theorem corrections, neutron-star and ISS timing applications.}}

\author{ Y. Friedman  and S. I. Klimovsky\\
Extended Relativity Research Center\\ The Applied Physics/Electro-Optics Engineering Department \\
Faculty for Engineering and Computer science \\   
Jerusalem College of Technology, Israel\\
P.O.B. 16031, Jerusalem 91160, Israel\\
e-mail: friedman@g.jct.ac.il}
\begin{document}
\date{}
\maketitle
\begin{abstract}
We investigate the gravitational field of an extended spherically symmetric body within the framework of Extended Relativity (ER), a Lorentz-covariant formulation of relativistic gravity on a Minkowski background. Using a relativistic superposition principle for retarded gravitational fields, we derive an explicit metric for an extended body by integrating the contributions of its mass elements.

The resulting metric reproduces the standard gravitational time dilation of a point source and agrees with the classical tests of General Relativity in the appropriate limits. However, unlike the exact Newtonian shell theorem and the Schwarzschild exterior solution, the external field depends weakly on the internal mass distribution through higher-order corrections. These corrections decay rapidly with distance but become significant near compact objects.

We analyze the corresponding admissible-velocity geometry, derive the motion equations for test particles, and compare the predictions of extended-body and point-source models. For neutron stars, the corrections noticeably modify the local light-velocity structure near the surface. For the Earth, the corrections are small but produce measurable differences in round-trip light travel times to the International Space Station.

The formalism provides a transparent relativistic description of extended gravitational sources and offers a framework for studying relativistic corrections due to internal structure in strong-field and precision-measurement regimes.

\end{abstract}

\section{Introduction}
In Newtonian gravity \cite{Newton}, the shell theorem states that a spherically symmetric mass distribution (such as a planet or star) acts as if all its mass were concentrated at a single point at its center, when viewed from an external point, and that the gravitational field at any point inside a spherically symmetric shell of mass is exactly zero (see, e.g., \cite[Note 3.1]{Kleppner}). However, is this theorem valid for relativistic gravitation?

\textit{General Relativity} (GR), formulated by Albert Einstein in 1915, is one of the most profound and revolutionary theories  of relativistic gravity \cite{Ein15}. For a comprehensive treatment of the foundations of $GR$, see Misner, Thorne, and Wheeler \cite{MTW}. A concise and classical presentation is given by Landau and Lifshitz \cite{Landau_Lifshitz}. Gr{\o}n provides a modern pedagogical introduction \cite{Gron}, while Kopeikin, Efroimsky, and Kaplan discuss applications in relativistic celestial mechanics \cite{Kop}. GR provides explanations for several phenomena that cannot be fully accounted for by Newtonian gravity, including the anomalous precession of Mercury’s orbit, the gravitational time dilation, the gravitational bending of light, gravitational waves, and others.

In GR, the classical notion of gravitation as a force is replaced by the curvature of space-time induced by mass-energy distributions. This is an implementation of Riemann's idea that \emph{``force equals geometry"}. The geometry is described by a metric, and the motion of objects corresponds to geodesics in this curved space-time. The metric itself is determined by Einstein's field equations from the distribution of energy and momentum. These equations form a system of nonlinear second-order differential equations. Solving this system is challenging, and due to its complexity, exact solutions exist only for specific cases. Throughout this work, we use the term \textit{gravitation} to refer to this curvature-based description.

In GR, the gravitational field outside a spherically symmetric, non-rotating mass distribution is described by the Schwarzschild metric \cite{Schw}. This solution is exact for a static black hole. A rotating black hole breaks spherical symmetry and is instead described by the axisymmetric Kerr metric \cite{Kerr}.
According to Birkhoff’s theorem \cite{Birk}, any spherically symmetric vacuum solution to Einstein’s equations must be static and asymptotically flat, and is given by the Schwarzschild solution, with a detailed modern exposition provided by Hawking and Ellis \cite{HE73}. Thus, the external gravitational field remains static even if the central mass distribution evolves dynamically, as long as spherical symmetry is preserved. This implies that outside a spherically symmetric mass distribution (even if the source is dynamic, e.g., pulsating), the spacetime is described by the Schwarzschild metric and depends only on the total mass.

Recent studies have revisited and extended these results. Lin \cite{Lin2024}, using a post-Newtonian treatment of collapsing thin shells, demonstrated how deviations from strict Birkhoff behavior can arise in dynamical scenarios that preserve spherical symmetry but are not static. Ovalle \cite{Ovalle2024} examined interior geometries compatible with the Schwarzschild exterior and showed how a complex extended source can be smoothly matched to it—an approach useful for modeling non–point-like interiors. Obukhov \cite{Obukhov2020} analyzed a generalized Birkhoff theorem within Poincaré gauge gravity and established the conditions under which static, spherically symmetric solutions persist in gauge-theoretic extensions of general relativity. Furthermore, Becerra, Ilijić, Narayan, and Rubio \cite{Becerra2024} investigated realistic anisotropic neutron stars and highlighted how pressure anisotropies within extended relativistic bodies modify the stellar interior and affect the matching to the external Schwarzschild spacetime, thereby underscoring the importance of extended-body effects in spherical configurations.

There are several approaches to extend the geometrization of physics also to the electromagnetic field. Misner and Wheeler \cite{Misner} proposed a model in which matter, charge, electromagnetism, and other fields are only manifestations of the bending of spacetime. The electromagnetic field emerges from the Ricci curvature tensor via the so-called “Maxwell square root” relation. Ludwig \cite{Lud70} re-derived parts of the above model using the Newman–Penrose  spin-coefficient formalism. Ferraris and  Kijowski \cite{Ferr} showed that electromagnetism can be embedded in spacetime geometry using connection-based formalisms instead of the metric. In addition, there are several approaches for unification of gravitation and electromagnetism in higher dimensions. The most well-known is due to Kaluza \cite{Kaluza} and Klein \cite{Klein}, who achieved the unification of these fields by adding a fifth dimension.

A new model for relativistic dynamics, known as Extended Relativity (ER), was first introduced in \cite{FUnify} and later developed in detail in the monograph \cite{ana}.
 This model unifies the dynamics of gravitational and electromagnetic fields without the need to add extra dimensions. ER successfully reproduces all known relativistic effects in both electrodynamics and general relativity. In this model, both electromagnetic and gravitational fields curve spacetime, but all dynamics are described in a flat spacetime background as observed by a distant observer, which simplifies the calculations. 

The model adopts Lorentz covariance instead of the general covariance employed in the works mentioned above. This choice is motivated by the need not only to satisfy the Principle of Relativity but also to achieve a unified geometric description of electromagnetic and gravitational fields. It also allows for simpler computations and enables superposition of fields from different sources. Extended Relativity relies on two new ideas:
\begin{enumerate}
    \item \textit{Influenced Spacetime} – Field sources curve spacetime. Each object experiences its own curved spacetime, determined by the fields acting upon it and its  charge-to-mass ratio.
    \item \textit{Extended Principle of Inertia} – An undisturbed object moves along the shortest (or stationary) worldline in its influenced spacetime. 
\end{enumerate}
This principle extends Newton’s First Law from free motion. In ER, every object is in free motion in its influenced spacetime.

To define the ``shortest" worldline, we must first define the ``length" of a worldline in the influenced spacetime. When projecting a curved globe onto a flat map, local distortions—scaling—are required at each point to accurately represent distances. As a result, a geodesic (the ``straight" line on the curved globe), such as an airplane trajectory, appears curved on the flat map. Similarly, we define a \textit{Local Scaling Function} ($LSF$) as the ratio of the interval between two nearby events $x$ and $x + \epsilon u$ in the influenced spacetime and the corresponding interval in flat spacetime. The $LSF$ is a function of a spacetime position $x$ and a four-vector direction $u$.

The $LSF$ must be Lorentz invariant to comply with the Principle of Relativity. Additionally, we require that $LSF$s be independent of the chosen evolution parameter. In \cite{ana}, the following simple $LSF$ is used to model electromagnetism, gravity, and the influence of a medium on light:
\begin{equation}\label{LSF}
    L(x,u) = \sqrt{(\eta_{\mu\nu}-h_{\mu\nu}(x)) u^\mu u^\nu} + kA_\mu(x) u^\mu,
\end{equation}
where $\eta_{\mu\nu}$ is the flat spacetime Minkowski metric, $A_\mu(x)$ is a Lorentz-covariant four-covector-valued function representing the electromagnetic field, and $h_{\mu\nu}(x)$ is a symmetric Lorentz-covariant $(0,2)$ tensor representing the gravitational field. The parameter \( k \) is the test object's charge-to-mass ratio.

The equation of motion for an object under gravitational and electromagnetic fields (i.e., stationary worldlines) is obtained by applying the Euler–Lagrange equations to the $LSF$.

The $LSF$ describes the geometry of an object’s influenced spacetime, extending the traditional notion of a metric. This generalization enables a unified geometric description of both the electromagnetic and gravitational fields. The current paper focuses solely on gravitational fields. In the absence of an electromagnetic field (\( A_\mu(x) = 0 \)), the square of the $LSF$ corresponds to a metric \( g_{\mu\nu}(x) \) defined by
\begin{equation}\label{metric decomposition}
    g_{\mu\nu}(x) = \eta_{\mu\nu} - h_{\mu\nu}(x) .
\end{equation}
More generally, any metric defined on Minkowski spacetime can be represented in this form, where $h_{\mu\nu}$ is a symmetric deviation tensor relative to the flat background metric $\eta_{\mu\nu}$.
Accordingly, gravitational dynamics can be described using the equivalent metric formulation, as in GR, and we adopt this approach throughout this paper.

For point-like gravitational sources, a new metric was defined \cite{FSuper}, constructed using the source’s retarded position and velocity. The formulation satisfies the Lorentz covariance, the Newtonian limit, and linearity of the \textit{deviation tensor} $h_{\mu\nu}$ with respect to the source mass, while providing an accurate description of the field near the horizon. This metric satisfies Einstein’s field equations and passes all standard tests of GR.

The standard formulation of electrodynamics begins with Maxwell’s equations and expresses the fields in terms of their sources. An alternative route—presented in Griffiths \cite{Griff}—is to first derive the Liénard–Wiechert retarded electromagnetic potential for a single moving point charge, and then, by applying the principle of superposition, to obtain the total potential for an arbitrary charge distribution. This approach naturally yields both the near-field and radiation-field components. In fact, most analyses of electromagnetic radiation rely on the use of retarded potentials, since they provide a clear and systematic way to separate the field into its near and far components.

ER adopts this second approach. For both the electromagnetic and gravitational fields, rather than starting from field equations, one begins by deriving the retarded potential of a single moving source for each field, and then obtains the total retarded potential by integrating over all sources. Since the single-source potential is a function of the retarded position and four-velocity of the source, for which the partial derivatives are known, one obtains explicit formulas for the acceleration caused by the field on a test object. Through this formulation,  the gravitational field can be meaningfully separated into near and far components: as in electrodynamics, the near field falls off as the square of the distance and depends on the relative position and four-velocity of the source, while the far field decays linearly with distance and depends also on the source's acceleration.

Since, in ER, the deviation from flat spacetime for a single source is proportional to the source mass, the approach allows for a consistent superposition of fields from multiple sources. This leads to a relativistic superposition principle for gravity within the ER framework, which will serve as the main tool used in this paper.

The continuous advancements in satellite technology and cosmological research have led to an increasing demand for precise descriptions of motion within gravitational fields, which necessitates the inclusion of relativistic corrections. Given that many gravitational sources exhibit approximate spherical symmetry, it is important to derive the relativistic corrections for motion within the gravitational field of an extended spherically symmetric source. In addition, the gravitational field holds information on the densities of matter inside the gravitating body. Knowing the connection between the sources and the field, we can learn from the accurate measurements of the field about the internal structure of the source.

Synge \cite{Synge} obtained a description of the gravitational field of an extended spherically symmetric body according to the gravitation theory of A. N. Whitehead \cite{White}. However, predictions of this theory contradict experimental facts, as explained in \cite{Gib}. In this paper, we derive an explicit metric of such a field, and obtain the relativistic corrections to motion in the field due to the size and non-uniform distribution of mass of the body.

In Section 2, we present the metric of a single-point source in Extended Relativity and describe the ball of admissible velocities in such a field. 
In Section 3, we apply the superposition principle to derive the metric of the gravitational field of a spherically symmetric body and compare it to that of a point-like source. In Section 4, we describe the ball of admissible velocities in such a field. We visualize this ball on the surface of a neutron star and describe its shape on the Earth's surface. We also compare the round-trip signal time delay between Earth and the International Space Station (ISS), considering both point-like and extended models of the Earth field. We found that the extended body correction is of the same order as the relativistic correction of a single point mass. In Section 5, we present a method to derive the equation of motion in the gravitational field of a spherically symmetric body.

%In Section 3, we introduce the superposition principle to construct the metric of a gravitational field generated by multiple sources. We also derive the equation of motion with respect to the coordinate time parameter in a general gravitational field. In Section 4, we obtain explicit formulas for the acceleration of objects in the gravitational field of a point-like, stationary source. We show that for circular orbits, the only relativistic correction arises from gravitational time dilation. Additionally, we demonstrate that light can follow a circular trajectory only on the photon sphere of a black hole.

\section{Metric of Single-Point Source in Extended Relativity } \label{ER model SPmetric}

Throughout this work, Greek indices $\mu, \nu, \alpha, \dots$ run over spacetime components $0,1,2,3$, while Latin indices $i, j, k, \dots$ refer exclusively to spatial components $1,2,3$.

In the framework of Extended Relativity (ER), the gravitational field is described by a metric tensor $g_{\mu\nu}(x)$ defined on the underlying Minkowski spacetime $(\mathbb{R}^4, \eta)$, as seen by a distant inertial observer. The background Minkowski metric is denoted by $\eta_{\mu\nu} = \text{diag}(1, -1, -1, -1)$, and all index raising and lowering in this work is performed using $\eta$. The motion of a test object in this field follows a geodesic with respect to the metric tensor $g_{\mu\nu}$. All geometric structures—such as light-cones, 1-forms, vectors, etc, are well-defined on the flat spacetime itself and not just in the tangent, cotangent spaces. This framework requires only \textit{Lorentz covariance}: all tensors must transform appropriately under Poincaré transformations of Minkowski space. General covariance is not assumed.

In ER, the formulation is based on global coordinates over Minkowski spacetime, where gravitational effects are modeled as deviations from flat geometry. This approach emphasizes Lorentz covariance and provides a clear and unified framework for describing both gravitational and electromagnetic fields.

Any metric tensor $g_{\mu\nu}(x)$ on Minkowski spacetime can be expressed using the flat spacetime metric $\eta_{\mu\nu}$ and a deviation from it $h_{\mu\nu}(x)$, as in \eqref{metric decomposition}. The \textit{deviation tensor} $h_{\mu\nu}(x)$ is 
 a symmetric tensor that encodes the gravitational field’s contribution to the distortion of spacetime geometry. The coordinates $x^\mu = (x^0, x^1, x^2, x^3)$ are the Minkowski space-time coordinates. Accordingly, the spacetime line element takes the form:
\begin{equation} \label{ds^2}
ds^2 = g_{\mu\nu}(x) dx^\mu dx^\nu = \left( \eta_{\mu\nu} - h_{\mu\nu}(x) \right) dx^\mu dx^\nu.
\end{equation}
%The minus sign in front of $h_{\mu\nu}(x)$ ensures that the gravitational field is attractive when $h_{\mu\nu}(x)$ is positive-definite.

As in the linearized theory of gravitation (see e.g., \cite{MTW}), the metric is decomposed with respect to $\eta_{\mu\nu}$. However, in the framework of ER, $h_{\mu\nu}(x)$ is not assumed to be small. Consequently, the decomposition in equation \eqref{metric decomposition} holds non-perturbatively and remains valid even in the presence of strong gravitational fields.

This approach differs from linear perturbation theory as used in cosmology. Mukhanov provides a detailed account of how gauge freedoms complicate the treatment of scalar, vector, and tensor modes \cite{Mukhanov}, while Weinberg discusses the resulting ambiguities in interpreting metric perturbations in a general relativistic framework \cite{WeinbergCosmo}. In contrast, the ER framework is formulated on a Minkowski background with Lorentz covariance only. The deviation tensor $h_{\mu\nu}$ has a direct physical meaning and is defined globally and covariantly, with no need for gauge fixing.

First, consider the gravitational field produced by a single point source of mass $m$, moving along a worldline $\psi(\tau)$. We express the mass in length units as $M = Gm/c^2$.

In General Relativity (GR), as manifested in the Schwarzschild solution, the deviation $h_{\mu\nu}(x)$ is a nonlinear function of the source's mass. This nonlinearity complicates the superposition of gravitational fields from multiple sources.

In the Extended Relativity (ER) framework, the deviation tensor $h_{\mu\nu}(x)$ is assumed to depend linearly on the source mass $M$. This linearity allows for the superposition of gravitational fields from multiple sources—an essential feature reminiscent of classical electromagnetism. Physically, it is reasonable to expect that the distortion of spacetime caused by a point mass should scale proportionally with the mass itself.

The deviation tensor $h_{\mu\nu}(x)$ is a symmetric tensor and can be interpreted as a symmetric matrix at each point. Since any symmetric matrix can be expressed as a linear combination of rank-one symmetric tensors of the form $l_\mu l_\nu$, for some covector $l_\mu$, it is natural to model the field of a single point-like mass using such a structure, assuming a single rank-one term rather than a sum.

Thus, ER postulates that the deviation tensor for a single point-like source takes the form
\begin{equation}\label{h_metric}
h_{\mu\nu}(x) = 2M\, l_\mu(x) l_\nu(x),
\end{equation}
where $l_\mu(x)$ is a Lorentz-covariant covector constructed from the observation point $x$ and the source's worldline $\psi(\tau)$.  We may refer to \( l_\mu(x) \) as the \textit{gravitational four-potential}. The prefactor $2M$ corresponds to the Schwarzschild radius  of the point source. 

This form of $h_{\mu\nu}$ is also a \emph{degenerate symmetric tensor} of Kerr–Schild type, introduced in \cite{KerrSc} and applied to several models in \cite{Adler}, — a structure known to yield exact solutions in general relativity and particularly well-suited for describing fields that admit linear superposition.

\begin{figure}[ht!]
\centering
\includegraphics[width=5 cm]{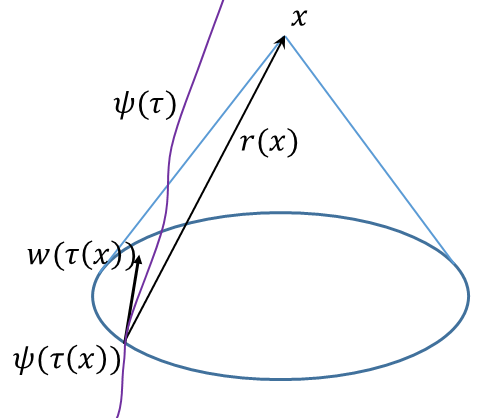}
\caption{ An object located at space-time point with coordinates $x$ in the gravitational field of a single moving source. The worldline of the source is denoted by $\psi(\tau)$.
The point $\psi(\tau(x))$ represents the unique intersection source’s worldline with the backward light cone $LC_x^-$ of the observation point $x$. $\tau(x)$ is the retarded time. The vector 
$r(x)$ connects the source’s position at the retarded time to the observation point.
The four-velocity of the source at the retarded time is $w(\tau(x))$.}
\label{Retardedpng}
\end{figure}

The explicit form of $l_\mu(x)$ is determined from Lorentz covariance and the Newtonian limit, as follows:
Define the \textit{retarded time} \(\tau= \tau(x) \)  as the unique solution to \( (x - \psi(\tau))^2 = 0 ,\) satisfying \( x^0 - \psi^0 (\tau(x)) > 0 \). The point with coordinates \( \psi (\tau(x)) \) is called the \textit{retarded position} of the source. This point is the unique intersection of the backward light cone with vertex at \( x \) and the source's worldline \( \psi (\tau) \), see Figure \ref{Retardedpng}.

Since the field propagates at the speed of light, only the source at the retarded time can influence the object at \( x \).
The four-velocity and four-acceleration of the source at the retarded time \( \tau(x) \) are denoted by \( w(x) = w(\tau(x)) \) and \( a(x) = a(\tau(x)) \), respectively. These four-vectors are Lorentz covariant. Another Lorentz covariant four-vector is the \textit{relative retarded position} of \( x \) with respect to the source's retarded position, given by \( r(x) = x - \psi(\tau(x)) \).

Since $l_\mu (x)$ must be Lorentz covariant, it can be constructed as a linear combination of the two four-vectors $r$ and $w$, with scalar functions of $r\cdot w$ as coefficients. Moreover, the acceleration of a slowly moving test particle, as determined by the geodesic equation, must reduce to the Newtonian gravitational acceleration in the appropriate limit.

For a single source, Kerr and Schild \cite{KerrSc} used a null covector $l_\mu$ that led them to the Schwarzchild metric in Eddington-Finkelstein coordinates \cite{Edding}. The problem with this solution is that the retarded Eddington–Finkelstein coordinates do not predict black holes, while the advanced Eddington–Finkelstein coordinates rely on advanced time, which contradicts causality.

Therefore, in \cite{FSuper},  a new covector $l_\mu$ was introduced, based on the retarded position, predict black holes and proper behavior near event horizon. To achieve this, we introduced a conjugation $^{*w}$  of the four-vector $r(x)$ with respect to the retarded velocity $w(x)$, defined by 
\begin{equation}\label{defConj}
    r_\mu^{*w}(x)=2(r(x)\cdot w(x))w_\mu(x)-r_\mu(x)\,.
\end{equation}
This conjugation reverses the spatial direction of the four-covector $r(x)$ in the frame moving with four-velocity $w$, while preserving its time component. 
If the direction of $r(x)$ corresponds to the propagation direction of the gravitational field at $x$ —whose spatial component points away from the source— then the spatial direction of $r^{*w}(x)$ is toward the source, like the gravitational  force. 
Using this notation, the covector $l_\mu(x)$ is given by,  
\begin{equation}\label{4potGmoving}
l_\mu(x)=\frac{r_\mu^{*w}(x)}{(r(x)\cdot w(x))^{3/2}}\,.
\end{equation}

This expression highlights the analogy with the electromagnetic four-potential; accordingly, we may refer to \( l_\mu(x) \) as the \textit{gravitational four-potential}.
For the case where the source is at rest, \( w = (1, 0, 0, 0) \). In this case, it is convenient to use $(1+3)$ decomposition of the background spacetime and to denote the spatial part of four-vectors in boldface. 
Second-rank tensors will, when appropriate, be expressed as block matrices corresponding to the \((1+3)\) decomposition, with time and spatial components explicitly separated, both for compactness and to emphasize the underlying spacetime structure.
Let $\mathbf{r}=(r^1,r^2,r^3)$ denote the spatial  vector from the source to the position of the object. Then the relative position null-vector is $r=(|\mathbf{r}|,\mathbf{r})$, and $r\cdot w= |\mathbf{r}|$. Since the co-vector $r_\mu=(|\mathbf{r}|,-\mathbf{r})$,  its conjugate is $r^{*w}_\mu=(|\mathbf{r}|,\mathbf{r})$,
and
\begin{equation}\label{4potGrest}
    l_\mu=\frac{(|\mathbf{r}|,\mathbf{r})}{|\mathbf{r}|^{3/2}}=\frac{(1,\mathbf{n})}{|\mathbf{r}|^{1/2}},
\end{equation} 
where $\mathbf{n}=\frac{\mathbf{r}}{|\mathbf{r}|}$ is the unit vector in direction of $\mathbf{r}$. 

Using \eqref{h_metric}, this implies that the deviation tensor of a single-source gravitational field, expressed as a block matrix, is
\begin{equation}\label{hStatSS}
  h_{\mu \nu} = \frac {2M} {|\mathbf{r}|}  \begin{pmatrix}
1 & n_i \\ n_j & n_i n_j
\end{pmatrix},
\end{equation}
where $M$ is the mass of the source, $\mathbf{r}$ is the vector from the source to the observation point, and $\mathbf{n}$ is its direction. 

When motion in a static single-source field is confined to a two-dimensional plane, it is convenient to use a $(1+2)$ decomposition, which we refer to as the \emph{time-polar} coordinates. In these coordinates, for any spacetime point, the first coordinate is the time component, the second is the radial spatial component from the source, and the third is the transverse spatial component
\begin{equation}\label{coordTimePolar}
   x=(x^0,r,r_\perp),\;\;\;  dx=(dx^0,dr,dr_\perp) \ .
\end{equation}
This coordinate system reflects the symmetry of the problem and simplifies the expressions by separating time, radial, and transverse components. We will adopt this coordinate choice frequently throughout the paper. In these coordinates, $\mathbf{n} = (1, 0)$, and from \eqref{hStatSS}, the deviation tensor is
\begin{equation} \label{h_single}
    h_{\mu\nu} = \phi \begin{pmatrix} 1 & 1 & 0 \\ 1 & 1 & 0 \\ 0 & 0 & 0 \end{pmatrix} ,
\end{equation}
where
\begin{equation}\label{phi}
    \phi(\rho)=\frac{2M}{|\mathbf{r}|}
\end{equation}
is a scalar function related to gravitational time dilation. Using \eqref{metric decomposition}, the metric of the gravitational field is
\begin{equation}\label{metricSSstat}
    ds^2=(1-\phi)(dx^0)^2-2\phi dx^0 dr-(1+\phi)dr^2-dr_\perp^2\,.
\end{equation}

In \cite{FStav}, this metric was shown to satisfy Einstein's field equations. It was also demonstrated that the geodesic motion of both massive and massless objects under this metric passes all the standard tests of $GR$.
%We may refer to the function \( l_\mu(x) \) as the\textit{ gravitational four-potential}.

The change in geometry due to a gravitational field can be observed through the characterization of admissible velocities (with respect to coordinate time $t$) for objects moving in the field. In the absence of a gravitational field, admissible velocities are bounded by the speed of light $c$. However, when the geometry is defined by the metric $g_{\mu\nu}(x)$ given in \eqref{ds^2}, the admissible four-velocities $u = dx / (c\, dt) = (1, \boldsymbol{\beta})$, where the unit-free spatial velocity is $\boldsymbol{\beta} = \mathbf{v} / c$, must satisfy the inequality $g_{\mu\nu}(x)\, u^\mu u^\nu \ge 0$.

For a single-source gravitational field, at any spacetime point $x$, the admissible four-velocities belong to the domain
\begin{equation}\label{AdmissFour_Vel}
 D_x=\{u:\;  1-\beta^2-2M(l_\mu(x)u^\mu)^2\geq 0\}
\end{equation}
for $l_\mu$ defined by \eqref{4potGmoving}. The velocity of light or any massless particles belong to the boundary of $D_x$,
 where $ds^2=0$. 

Consider now a gravitational field of a static black hole of mass $M$, where the entire mass is within the Schwarzchild radius $2M$. This field can be considered as a field generated by a rest-point source, for which $l_\mu$ is defined by \eqref{4potGrest}.
At each  spatial distance $\rho$ from the black hole the field limits the ball of admissible (3D) velocities at this point. 
We illustrate the 2D section of this ball and its boundary - the light velocities $\partial D_x$ at different spatial distances $\rho$ from the black hole. We use time-polar coordinates \eqref{coordTimePolar}. Decompose the unit-free velocity $\mathbf{\boldsymbol{\beta}}$ of light  in the field into radial $\beta_r$ and transverse $\beta_\perp$ components. The four-velocity  is $u=(1,\beta_r,\beta_\perp)$. This implies that $(l(\rho)_\mu  u^\mu)^2=(1+\beta_r)^2/\rho$. Using the definition \eqref{phi} of $\phi$, the admissible velocities must satisfy
\[ 1-\beta^2-\phi(1+\beta_r)^2\ge0,\]
or
\begin{equation}\label{BallGrav}
(1+\phi)\beta_r^2 +2\phi\beta_r+\beta_\perp^2\le1-\phi.
\end{equation}
The canonical form of the velocity-of-light ellipsoid equation is
\begin{equation}
    (1+\phi)^2\left(1+\beta_r-\frac{1}{1+\phi}\right)^2+(1+\phi)\beta_\perp^2=1.
\end{equation}
Note that the maximum value of $\beta_\perp$ is $\frac{1}{\sqrt{1+\phi}}$ and if the velocity of a massless particle at $\rho$ is transverse, then 
\begin{equation}\label{LightTransVel}
\beta_\perp^2(\rho)=1-\phi(\rho) .
\end{equation} 

\begin{figure}[!htb]
\centering
 \scalebox{0.4}{\includegraphics{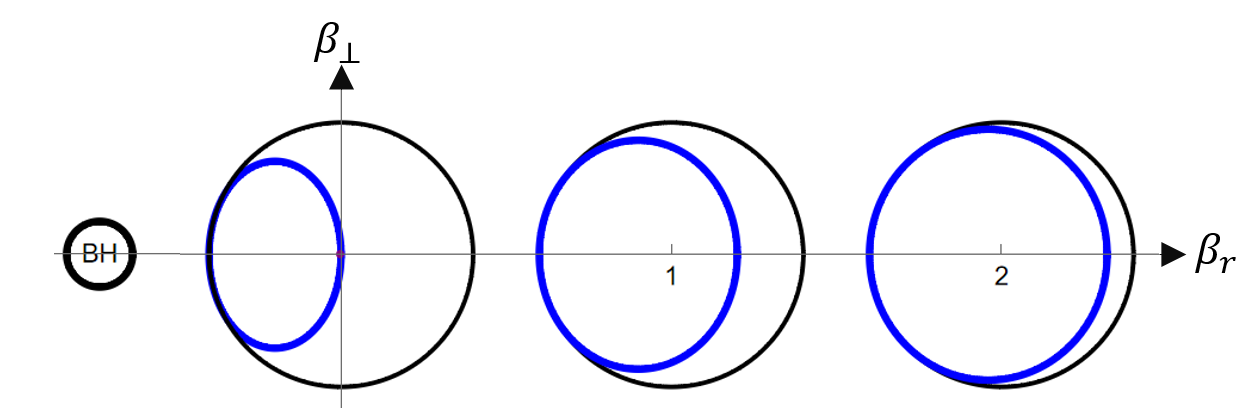}}
\caption{The velocity of light in the vicinity of a black hole. The figure shows the 2D section of the velocity of light in flat spacetime (black) and influenced spacetime (blue), at distances $\rho = 2M$, $6M$, and $18M$ from the black hole. }
\label{BH_new}
\end{figure}

The boundary of inequality (\ref{BallGrav}), is an ellipse in $(\beta_r,\beta_\perp)$ of admissible velocities of light. Direct substitution shows that $\beta_r=-1,\beta_\perp=0$ satisfies this equation. This means that in the direction $-\mathbf{r}$ towards the source, the speed of light is $c$, the speed of light in vacuum. This means that the light in the direction of the field will not be slowed down by the field. Another solution is 
\begin{equation}\label{velRadlight}
    \beta_r=\frac{1-\phi}{1+\phi}, \ \beta_\perp=0,
\end{equation}
which means that in the direction $\mathbf{n}$ away from the source, the speed of light is reduced by the factor $\frac{1-\phi}{1+\phi}$. At the Schwarzschild radius, where $\phi=1$, the speed of light becomes zero, showing that even light cannot escape the region inside the Schwarzschild radius. See Figure \ref{BH_new}

\section{The Metric of the Gravitational Field of a Spherically Symmetric Body}

In GR, the superposition of gravitational fields from multiple sources is typically achieved using post-Newtonian \cite{Kop1}, or post-linear approximations  \cite{Bel}.
Since the deviation tensor $h_{\mu\nu}$, as defined in \eqref{h_metric}, is linearly proportional to the mass of the source, we may assume that the deviation tensor for a field generated by several sources $M_j$ is given by the sum of the individual contributions from each source \cite{FSuper}:
\begin{equation}\label{devComb}
    h_{\mu\nu}(x) = \sum_j 2M_j\, l_{j\mu}(x)\, l_{j\nu}(x) \,,
\end{equation}
where $j$ labels the sources, and $l_{j\mu}(x)$ denotes the $\mu$-th component of the gravitational four-potential associated with the $j$-th source.
The corresponding metric for the combined field is then obtained using \eqref{metric decomposition}.

To compute the gravitational field of a spherically symmetric body, we first describe the mass distribution geometrically in 3D space ($\mathbb{R}^3$), providing intuitive clarity. This description is performed in a global inertial frame in which the body is at rest. The full computation of the gravitational metric is then carried out in Minkowski spacetime ($\mathbb{R}^4$), using the extended relativity framework. Once the metric is obtained, the analysis is carried out in the time-polar coordinates defined in equation \eqref{coordTimePolar}, which are chosen to reflect the symmetry of the problem and to simplify the resulting expressions.

Consider the gravitational field generated by a spherically symmetric body — a ball $B$ of radius $R$ and mass $M$ (in length units) at rest. We compute the field metric at a point $P$ outside the ball, at a distance $\rho$ from its center, with $\rho > R$.  
Due to the spherical symmetry of the problem, we may assume that the point $P$ lies on the $z$-axis. This choice makes the $z$-axis the radial direction, and the $x$- and $y$-axes the transverse directions. The metric of the gravitational field of the ball is derived from the deviation tensor $h_{\mu\nu}$.

To calculate the deviation tensor $h_{\mu\nu}$, we first decompose the ball $B$ into spherical shells, and then decompose each shell into differential mass elements. The deviation tensor of a single shell, $h'_{\mu\nu}$, is obtained by superposing (integrating) the corresponding deviation tensors of the differential mass elements within the shell. The deviation tensor of the entire ball, $h_{\mu\nu}$, is then obtained by integrating the deviation tensors of the individual shells.

\begin{figure}[ht]
    \centering
    \includegraphics[width=8cm]{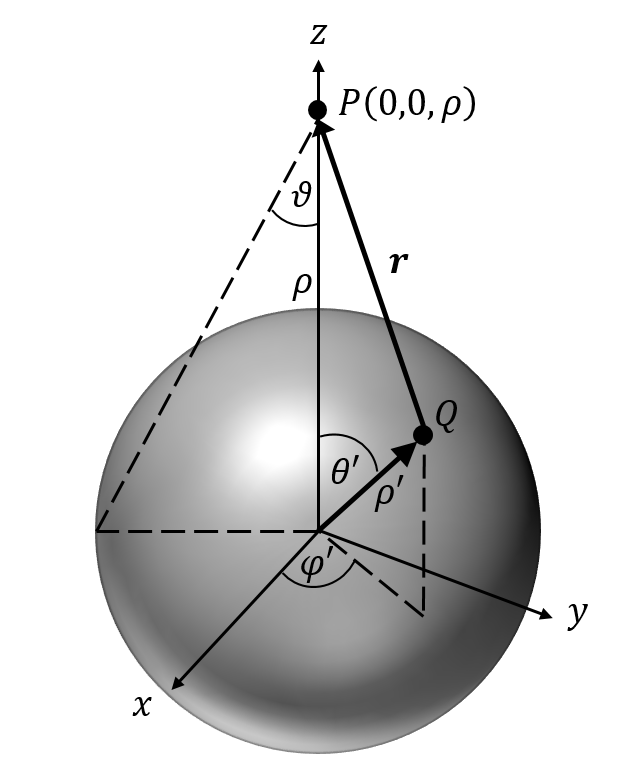}
    \caption{A differential mass element on a spherical shell of radius $\rho'$, is positioned at $Q(\rho',\theta',\varphi')$.  The observation point $P(0,0,\rho)$ of the field, outside the shell, is on the $z$ axis. The vector $\mathbf{r}$ denotes the relative position from $Q$ to $P$. The angle of perspective of the shell from $P$ is $\vartheta$.}
    \label{fig:shell}
\end{figure}

Consider a spherical shell of mass $M'$ and radius $\rho'$. Let $Q(\rho', \theta', \varphi')$ denote the position of a differential mass element on the shell, described in spherical coordinates, see Figure \ref{fig:shell}.

The Cartesian coordinates of $Q$ are
$$
 \rho' (\sin\theta' \cos\varphi',\; \sin\theta' \sin\varphi',\; \cos\theta')
$$
and the observation point $P$ is located at $ (0, 0, \rho)$. The relative position vector $\mathbf{r}$, pointing from the source point $Q$ to the observer at $P$, is:
$$
\mathbf{r} = 
(-\rho' \sin\theta' \cos\varphi',\; -\rho' \sin\theta' \sin\varphi',\; \rho - \rho' \cos\theta').
$$

Introducing the dimensionless parameter
\begin{equation}\label{q_def}
q = \frac{\rho'}{\rho}, \quad \text{with } 0 < q < 1,
\end{equation}
which characterizes the size of the shell relative to the observation point, we can express $\mathbf{r}$ as:
$$
\mathbf{r} = \rho \left(-q \sin\theta' \cos\varphi',\; -q \sin\theta' \sin\varphi',\; 1 - q \cos\theta'\right).
$$

The parameter $q$ is also related to the perspective angle $\vartheta$ of the shell as seen from the point of view of the observer, through the relation $q = \tan \vartheta$. This parameter expresses the difference between the field of a point source (for which $q=0$) and that of an extended source.

The magnitude of the vector $\mathbf{r}/\rho$ is given by:
\begin{equation}
g(q, \theta') = |\mathbf{r}|/\rho = \sqrt{q^2 + 1 - 2q \cos\theta'},
\end{equation}
and the direction of the relative position is
\begin{equation}
   \mathbf{n} = \frac{\mathbf{r}}{|\mathbf{r}|}= \left(- \frac{q}{g}\sin\theta' \cos \varphi' \ ,  \ -\frac{q}{g}\sin\theta' \sin \varphi' \ , \ \frac{1}{g} -\frac{q}{g}\cos\theta' \right) .
\end{equation}

Assuming that the mass is uniformly distributed over a single shell, decompose the mass $M'$ into differential masses given by $dM'=\frac{M'}{4\pi \rho'^2}dS' =\frac{M'}{4\pi} \sin\theta' d\theta' d\varphi'$. 
Using  \eqref{devComb} and \eqref{hStatSS}, the deviation tensor $ h'_{\mu\nu}$ of the field of the shell at $P$ is given by:
\begin{equation} \label{H_integral_equation}
    h'_{\mu\nu} = \frac{2M'}{4\pi} \int_{0}^{2\pi} \int_{0}^{\pi}  \frac {1} {\rho g(q,\theta')} \cdot \begin{pmatrix}
1 & n_i \\ n_j & n_i n_j
\end{pmatrix}\sin\theta'  d\theta' d\varphi'\,.
\end{equation}

Since the integrals of $\sin\varphi'$ and $\cos\varphi'$ from $0$ to $2\pi$ are zero, most components in the tensor vanish. The non-zero diagonal components are (see Appendix I) $h'_{00}=\frac{2M'}{\rho}$, $h'_{11}=h'_{22}=\frac{2M'q^2}{3\rho }$ and $h'_{33}=\frac{2M'}{\rho}\left(1-\frac{2q^2}{3}\right)$. The non-zero off-diagonal components are
\[h'_{03}=h'_{30}=\frac{M'}{\rho}\left(\frac{1}{2}(1-q^2)\ln{\frac{1+q}{1-q}}  +  q \right)=\frac{2M'}{\rho}\left(1-\frac{q^2}{3}-b(q)\right) ,\]
where
\begin{equation}\label{b(q)def}
 b(q)=\sum_{n=2}^{\infty} \frac{q^{2n}}{(2n-1)(2n+1)}.\end{equation}
 Thus, the deviation tensor for a spherical shell of radius \( \rho' \) and mass \( M' \), assuming a uniform mass distribution, is given by:
\begin{equation}\label{hSS}
     h'_{\mu\nu} = \frac{2M'(\rho')}{\rho} 
\begin{pmatrix}
1 & 0 & 0 &1- \frac{q^2}{3}-b(q)
\\ 0 & \frac{q^2}{3} & 0 & 0 \\
0 & 0 & \frac{q^2}{3} & 0 \\
1-\frac{q^2}{3}-b(q) & 0 & 0 &  1 - \frac{2q^2}{3}
\end{pmatrix}
\end{equation}

To obtain the deviation tensor for the entire ball $B$ of radius $R$ and mass density $\sigma_m(\rho')$, we divide the ball into spherical shells. The differential mass of a shell between radius $\rho'$ and $\rho'+d\rho'$ is $dM'(\rho ')=\sigma_m(\rho')  4\pi\rho'^2 d\rho'$. 

The mass moments $M_n$ describing the mass distribution in the ball are defined by:
\begin{equation} \label{Mn}
    M_n = 4\pi \int_0^R \sigma_m(\rho')  \rho'^n d\rho' \quad (M_2=M)
\end{equation}
For a uniform mass distribution,
\begin{equation}
    M_n = 4\pi \int_0^R \frac{M}{4\pi R^3/3}  \rho'^n d\rho' = \frac{3MR^{n-2}}{n+1} .
\end{equation}
Then, the deviation tensor of the ball is
\begin{equation}\label{devFrom Shell}
    h_{\mu\nu}=\int_0^R h'_{\mu\nu}(\rho ')d\rho'\,.
\end{equation}
As seen from \eqref{hSS}, all components of this tensor consist of terms proportional to $M'(\rho') q^k$. Therefore, computing the full deviation tensor reduces to evaluating integrals involving these expressions.
\begin{equation} \label{integral_Mq^k}
    \int_0^R dM'(\rho')q^k=\int_0^R \sigma_m(\rho')  4\pi\rho'^2\frac{(\rho')^k}{\rho^k} d\rho'=\frac{M_{k+2}}{\rho^k} .
\end{equation}
Integrating over the spherical shells using this identity yields the total deviation tensor of the ball $B$.

\begin{equation} \label{h_ball}
   h_{\mu\nu} = \phi
\begin{pmatrix}
1 & 0 & 0 & 1
\\ 0 & 0 & 0 & 0 \\
0 & 0 & 0 & 0 \\
1 & 0 & 0 & 1
\end{pmatrix} 
+
\begin{pmatrix} 
0 & 0 & 0 & -\psi-\chi
\\ 0 & \psi & 0 & 0 \\
0 & 0 & \psi & 0 \\
-\psi-\chi & 0 & 0 & -2\psi
\end{pmatrix}
\end{equation}
where $\phi(\rho)= \frac{2M}{\rho}$ is defined by \eqref{phi},
\begin{equation} \label{psi}
    \psi = \frac{2M_4}{3\rho^3}
\end{equation}
\begin{equation} \label{chi}
    \chi = \chi(\rho) = \sum_{n=2}^{\infty} \frac{2 M_{2n+2}}{(2n-1)(2n+1) \rho^{2n+1}} \, .
\end{equation}
The first matrix corresponds to the deviation tensor of a point-like source, while the second captures the contribution of the extended mass distribution. The dominant correction due to the extended nature of the source is given by $\psi$, while higher-order corrections are represented by $\chi$.

For a uniform mass distribution:
\begin{equation}\label{psichi_uni}
\psi = \frac{2M}{\rho} \frac{(R/\rho)^2}{5} \ , \quad \chi = \frac{2M}{\rho} \sum_{n=2}^{\infty} \frac{3(R/\rho)^{2n}}{(2n-1)(2n+1)(2n+3)} .
\end{equation}
In this case, the deviation tensor is linear in the total mass $M$ of the ball. However, it depends not only on the distance $\rho$ from the observation point to the center of the ball, but also on the ratio $\frac{R}{\rho}$ between the ball’s radius and this distance.

To understand the metric of the gravitational field outside the ball $B$ at an arbitrary point $P$, we use the time-polar coordinates defined in \eqref{coordTimePolar}. In these coordinates, the matrix representation of the metric is:
\begin{equation}\label{metricBallMatrix}
    g_{\mu\nu} = \
\begin{pmatrix}
1-\phi & -\phi +\psi+\chi & 0   \\
-\phi+\psi+\chi & -1-\phi+2\psi & 0 \\0& 0& -1-\psi
\end{pmatrix}
\end{equation}
and the corresponding line element is:
\begin{equation}\label{metricBall}
ds^2 = (1 - \phi)(dx^0)^2 - 2(\phi - \psi - \chi) dx^0 dr - (1 + \phi - 2\psi)dr^2 - (1 + \psi)dr_{\perp}^2.
\end{equation}

Compare this metric with that of a point-like source given in \eqref{metricSSstat}. We first observe that the gravitational time dilation—determined by the $g_{00}$ component—is identical in both the extended body and point-source cases. This mirrors the Newtonian shell theorem. However, the metric corresponding to an extended mass distribution includes additional terms involving the potentials $\psi$ and $\chi$, which are smaller in magnitude than the leading potential $\phi$ associated with the point source.

In the far-field region, where $\rho \gg R$, the terms decay at different rates: $\phi$ falls off as $1/\rho$, $\psi$ as $1/\rho^3$, and $\chi$ as $1/\rho^5$. Therefore, the metric of the extended body field asymptotically approaches that of the point-like source. While the Newtonian shell theorem does not hold exactly in the ER model, it remains a good approximation in the far-field regime.

\section{Admissible Velocity Ball in an Extended Body Gravitational Field}

As for the field of a point-like source, also for the field of an extended body, we will look for the effect of the field on the domain of admissible velocities. In such a field, this domain will also depend on the position of the moving object, mainly on its distance from the center of the source. Due to spherical symmetry, it is sufficient to solve the problem within the subspace spanned by the radial direction to the position of the object and the velocity of the object. The radial direction will be denoted by \( x^1 \), and the transverse direction by \( x^2 \). 

To define the surface of admissible light velocities at a point at distance $\rho$ from the center of the source, we decompose the velocity of the object into a radial component $\beta_r$ and a transverse component $\beta_\perp$. Since for light, \( ds = 0 \), from \eqref{metricBall} we obtain :
\begin{equation} 
    (1-\phi) = 2(\phi-\psi-\chi) \beta_r + (1+\phi-2\psi)\beta_r^2 + \left(1+\psi \right) \beta_\perp^2 .
\end{equation} 
This surface is the boundary of the admissible velocity ball in an extended body gravitational field.

Denote
\begin{equation}
    a=1+\phi-2\psi \ , \ b = \phi-\psi-\chi \ , \ c = 1+\psi  \ , \ d = 1-\phi .
\end{equation}
With this notation, the  admissible light velocities ellipsoid surface in canonical form becomes
\begin{equation}
     \frac{a}{d +\frac{b^2}{a}}(\beta_r + \frac{b}{a})^2 +\frac{c}{d +\frac{b^2}{a}}\beta_\perp^2 = 1 .
\end{equation}
Introducing some parameter $\theta$ from  $[0,2\pi]$, this surface can be defined as:
\begin{equation} \label{velocity_vall_extended_eq}
    \beta_r = \sqrt{\frac{d +\frac{b^2}{a}}{a}}\cos\theta - \frac{b}{a}  \ , \;
\beta_\perp = \sqrt{\frac{d +\frac{b^2}{a}}{c}}\sin\theta \,.\end{equation}
For a single point source field such surface is
\begin{equation}\label{velocity_bal_sp}
   \beta_r = \frac{\cos\theta}{1+\phi} - \frac{\phi}{1+\phi} \ , \;\ \beta_\perp =\frac{\sin\theta}{\sqrt{1+\phi}}. 
\end{equation}

\subsection{Velocity ball on a neutron star surface}\label{neutron}

We  examine first the effect of the extended body corrections on the admissible velocity sphere for a neutron star.  As noted in \cite{Neutron}, neutron stars typically have a mass of approximately 1.5 solar masses, corresponding to a Schwarzschild radius of about 4 km, with an actual radius \(R\) of around 12 km. For a typical neutron star with a radius \(R = 3r_s = 6M\), we will illustrate the admissible velocity sphere at its surface, i.e., at \(\rho = R\), and compare it to that of a point mass source. For simplicity, we assume a uniform mass distribution, where the mass moments are given by \( M_n = \frac{3MR^{n-2}}{n+1} \). Then,
from equations \eqref{phi} and \eqref{psichi_uni}
\begin{equation}
   \phi=\frac{1}{3} \ \ , \ \ \psi =\frac{1}{15} \ \ ,\ \  \chi= \frac{1}{60} .
\end{equation}
Thus, the admissible light velocity ellipsoid on neutron star surface, defined by \eqref{velocity_vall_extended_eq}, is
\[ \beta_r = 0.7734\cos\theta - 0.2083 ,\,\; \beta_\perp = 0.821\sin\theta \]
The similar ellipsoid, if we would consider the star as a single point, using \eqref{velocity_bal_sp} is
\[ \beta_r = 0.75\cos\theta - 0.25,\;\;\beta_\perp = 0.866\sin\theta \]

\begin{figure}[ht]
    \centering
    \includegraphics[width=8cm]{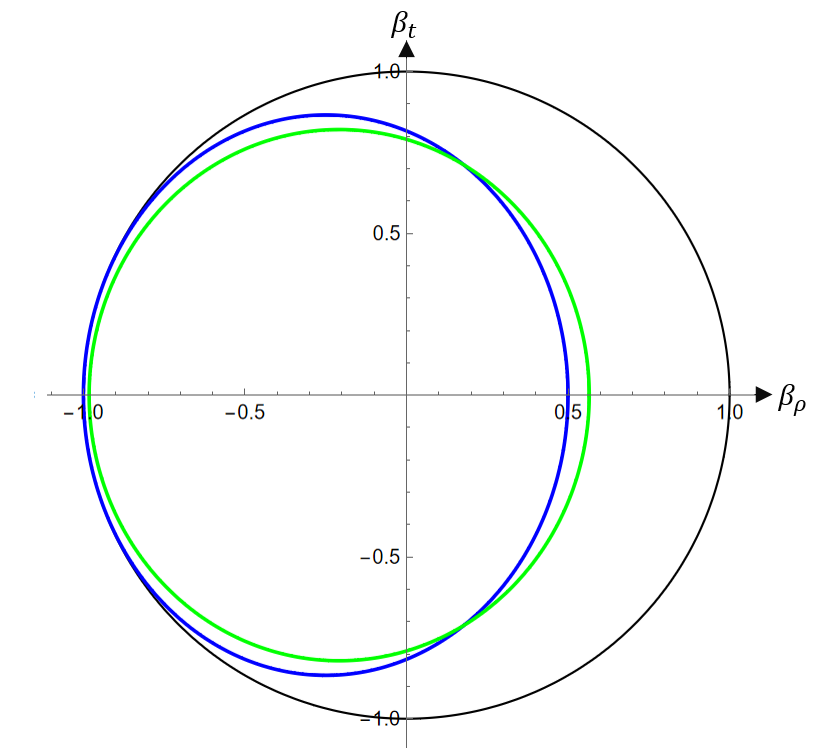}
    \caption{The admissible light velocity ellipsoid 2D section on  the surface of a neutron star: (1) without gravitational field corrections (black), (2) in gravitational field of a single point source (blue), (3) in the gravitational field as an extended body (green).}
    \label{fig:velocity ball extended body}
\end{figure}

In Figure \ref{fig:velocity ball extended body} we present the admissible light velocity ellipsoid  on neutron star surface for both the $ER$ extended body model and as a single point source. 

The figure shows that, for an extended body, the admissible velocity in the direction outward from the source is greater than that for a single point source, while it is smaller in the transverse direction. In the direction toward the source, for a single point source the velocity of light is the speed of light in vacuum $c$, while for an extended body this velocity becomes slightly less than $c$. This occurs because for an extended body, the acceleration from each mass element is not exactly in the direction to the center of the body.
Since the gravitational field of an extended body is kind of averaging of the field from each mass element, the ellipse becomes closer to a circle. 

These corrections exist for any extended body, but they become particularly significant for very massive bodies, such as neutron stars, and only in the regions close to them.

\subsection{Velocity ball on the Earth surface}\label{Earth}
Now, we will examine the effect of the extended body corrections on the admissible velocity in the case of a weak gravitational field, such as that of Earth, and compare it to that of a point-mass source. Using these results, we can determine the velocity of light moving outward and inward relative to Earth.
For a weak gravitational field region $2M<<\rho$ we get $\frac{b^2}{a}<<d$ therefore
\begin{equation} \label{velocity_vall_extended_eq_Weak}
     \beta_r = \sqrt{\frac{d}{a}}\cos\theta - \frac{b}{a},\;\;\beta_\perp = \sqrt{\frac{d}{c}}\sin\theta  .
\end{equation}
For Earth, \(R = 1.44 \cdot 10^9 M\). For simplicity, we assume a uniform mass distribution. Using \eqref{psichi_uni}, close to Earth's surface, where $\rho\approx R$, we get
\begin{equation}
 \phi=1.3\cdot 10^{-9} \ \ , \ \ \psi =2.78\cdot 10^{-10} \ \ ,    \chi= 6.94 \cdot 10^{-11} .
\end{equation}
%Therefore
%\[a= 1.2 \ , \ b = 0.25 \ , \ c = \frac{16}{15} \ , \ d = \frac{2}{3} .\]
Thus,
\[ \beta_r = (1-1.114 \cdot10^{-9})\cos\theta - 1.044 \cdot10^{-9},\;\;\beta_\perp = (1-8.353 \cdot 10^{-10})\sin\theta \]
If we use the single point for the gravitational field, we obtain
\[\beta_r = (1-1.392 \cdot10^{-9})\cos\theta - 1.392 \cdot10^{-9},\;\;\beta_\perp = (1-6.961 \cdot 10^{-10})\sin\theta  \]

When treating the gravitational field of the Earth as an extended body source, the speed of light in the outward radial direction is \(\beta_r = 1 - 2.158 \cdot 10^{-9}\), and in the inward direction it is \(\beta_r = -1 + 6.96 \cdot 10^{-11}\). In contrast, when treating Earth as a single point gravitational source, the speed of light in the outward radial direction is \(\beta_r = 1 - 2.784 \cdot 10^{-9}\), and in the inward direction it is \(\beta_r = -1\). There is a 20\% difference in the relativistic corrections between both cases in the outward direction. In the inward direction, the deviation between the relativistic corrections is much smaller.

\subsection{Round-Trip Time Delay from Earth to the ISS}\label{roundtrip}

Accurately measuring the one-way speed of light is challenging due to the need for synchronized clocks at both ends of the path. In contrast, the round-trip (two-way) travel time of light can be measured precisely using a single clock at the point of emission and reception. This measurement allows us to examine how a gravitational field affects the round-trip travel time of light.

For a point mass, both the Schwarzschild and $ER$ metrics predict the same round-trip time for light propagation \cite{FStav}. However, as illustrated in Figure \ref{fig:velocity ball extended body}, when the gravitational field is generated by an extended body rather than a point source,  the predicted round-trip time differs slightly.

Consider, for example, the time it takes for light to travel from Earth to the International Space Station (ISS), which orbits approximately 400 km above Earth's surface. Neglecting gravitational effects, the expected round-trip time is $\Delta t = \frac{2 \times 400\, \text{km}}{c} \approx 2.67\, \text{ms}$.

In the presence of a gravitational field, the effective speed of light is reduced compared to its value in vacuum, leading to an increase in round-trip time. Both the Schwarzschild and $ER$ models, when treating Earth as a point mass, predict an increase of approximately 3.6 picoseconds (see Appendix II).

However, if we instead model Earth as an extended, spherically symmetric body with a uniform mass distribution, the predicted increase in travel time becomes asymmetric. Light traveling from Earth to the ISS experiences an increase of 2.85 picoseconds, while light traveling from the ISS back to Earth sees an increase of only 0.075 picoseconds. The total round-trip delay in this extended body model is therefore approximately 0.7 picoseconds shorter than that predicted by the point-mass models.

This demonstrates that even small relativistic corrections, such as the gravitational time delay, are sensitive to the mass distribution of the source. Treating Earth as an extended body leads to a measurably different prediction for relativistic corrections to the round-trip time of light.

\section{Acceleration in a Gravitational Field of an Extended Body}

The motion of objects in a gravitational field is described by geodesics in the influenced spacetime. This motion is governed by the geodesic equation. To facilitate its computation within the framework of Extended Relativity, we introduce the \textit{gravitational tensor} $G^{\alpha}_{\mu\nu}$, defined by raising the first index of the Christoffel symbols of the first kind constructed from the deviation tensor $h_{\mu\nu}$:
\begin{equation} \label{G_tensor}
G^{\alpha}_{\mu\nu} \coloneqq \frac{1}{2} \eta^{\alpha\beta} (h_{\beta\mu,\nu} + h_{\beta\nu,\mu} - h_{\mu\nu,\beta}).
\end{equation}
The gravitational tensor $G^\alpha_{\mu\nu}$ is linear in the mass distribution of the sources. For a single point-source field, using the explicit form of $h_{\mu\nu}$ defined by \eqref{h_metric} and \eqref{4potGmoving}, and the known expressions for the partial derivatives of the retarded position and retarded four-velocity \cite{Jackson}, an explicit form of the gravitational tensor was derived in \cite{FSuper}.

The Christoffel symbols $\Gamma^{\alpha}_{\mu\nu}$ can be expressed as
\begin{equation} \label{Christoffel}
\Gamma^{\alpha}_{\mu\nu} = -\left( (I-H)^{-1} \right)^\alpha_{\ \lambda} G^\lambda_{\mu\nu},
\end{equation}
where $I^\beta_\mu$ is the identity operator, and the \textit{deviation operator} $H^\beta_\mu$ is defined by
\begin{equation}\label{H_operator}
H^\beta_\mu = \eta^{\beta\nu} h_{\nu\mu}.
\end{equation}

Using these definitions, the acceleration of an object with respect to the arc-length parameter $s$ (defined via the line element \eqref{ds^2}) in the combined gravitational field is given by
\begin{equation} \label{acceleration_s}
\frac{d^2x^\alpha}{ds^2} = \left( (I-H)^{-1} \right)^\alpha_{\ \lambda} G^{\lambda}_{\mu\nu} \frac{dx^\mu}{ds} \frac{dx^\nu}{ds}.
\end{equation}

Although \eqref{acceleration_s} has a simple form, the parameter $s$ depends on both the position and velocity of the object. For practical purposes, especially when comparing with observations, it is more convenient to use the coordinate time $t = x^0$ as the evolution parameter, since it is independent of the object’s state. We work in units where $c = 1$.

The detailed transformation from $s$ to $t$ is given in Appendix III.
We denote derivatives with respect to the coordinate time $t$ by a dot, e.g., $\dot{x}^\mu = \frac{dx^\mu}{dt}$. Using this notation, we define the \textit{quadratic four-acceleration} as
\begin{equation} \label{quadratic four-acceleration}
\ddot{x}^\alpha_{[2]} = \left( (I - H)^{-1} \right)^\alpha_{\ \lambda} G^\lambda_{\mu\nu} \dot{x}^\mu \dot{x}^\nu.
\end{equation}
Then, the equation of motion takes the form
\begin{equation} \label{acceleration_tau}
\ddot{x}^\alpha = \ddot{x}^\alpha_{[2]} - \ddot{x}^0_{[2]} \dot{x}^\alpha,
\end{equation}
where $\ddot{x}^0_{[2]}$ accounts for the transformation between the arc-length parameter $s$ and the time parameter $t$. The last term ensures that $\ddot{x}^0 = 0$ in a frame comoving with the object, expressing the fact that the coordinate time is independent of position within the field.

This formulation is particularly advantageous because both the deviation operator $H$ and the gravitational tensor $G$ for a single point source are linearly proportional to the source's mass. Hence, we may assume that in the case of a gravitational field generated by multiple sources $M_j$, both $H$ and $G$ can be obtained by summing the individual contributions from each source. Once the combined tensors are calculated, they can be directly substituted into the equation of motion \eqref{acceleration_tau} to describe the object's trajectory in the total gravitational field.

In \cite{FSuper}, an explicit expression for the tensor $G^\alpha_{\mu\nu}$ of a single source field, defined by \eqref{G_tensor}, \eqref{h_metric} and \eqref{4potGmoving}, was derived. Using the known derivatives \cite{Jackson} of the retarded position and four-velocity vectors, this tensor is a function of these vectors.
We use the standard notation $x_{(\mu} y_{\nu)}$, defined for any two covectors $x$ and $y$ as:
\begin{equation} \label{vee_product}
    x_{(\mu} y_{\nu)} \coloneqq \frac{1}{2} (x_\mu y_\nu + x_\nu y_\mu) .
\end{equation}
The resulting expression for the gravitational tensor of a non-accelerating point source is:
\[G^{\alpha}_{\mu\nu} = M \left( \frac{2\eta_{\mu\nu}}{(r \cdot w)^3} - \frac{6w_{(\mu} r_{\nu)}}{(r \cdot w)^4} + \frac{3r_{(\mu} r_{\nu)}}{(r \cdot w)^5} \right) r^\alpha \]
\begin{equation} \label{G_near}
    + M \left( \frac{-4(\eta_{\mu\nu} + w_{(\mu} w_{\nu)})}{(r \cdot w)^2} + \frac{16w_{(\mu} r_{\nu)}}{(r \cdot w)^3} - \frac{7r_{(\mu} r_{\nu)}}{(r \cdot w)^4} \right) w^\alpha,
\end{equation}

For a source at rest, $w=(1,0,0,0)$, we have $r\cdot w = r^0$. Introducing $n^\alpha = \frac{r^\alpha}{r^0} = (1,\mathbf{n})$, we can rewrite \eqref{G_near} as
\[G^{\alpha}_{\mu\nu} = \frac{M}{r_0^2} \left( 2\eta_{\mu\nu} - 6w_{(\mu} n_{\nu)} + 3 n_{(\mu} n_{\nu)} \right) n^\alpha \]
\begin{equation}
    + \frac{M}{r_0^2} \left( -4(\eta_{\mu\nu} +w_{(\mu} w_{\nu)}) + 16w_{(\mu} n_{\nu)} - 7 n_{(\mu} n_{\nu)} \right) w^\alpha,
\end{equation}
Separating the time and spatial components, using Latin indices $i,j,k \in \{1,2,3\}$, we obtain
\[G^{0}_{\mu\nu} = \frac{M}{r_0^2} \left( -2\eta_{\mu\nu} -4w_{(\mu} w_{\nu)} + 10w_{(\mu} n_{\nu)} -4 n_{(\mu} n_{\nu)} \right) \]
\begin{equation}
   G^k_{\mu\nu} =  \frac{M}{r_0^2} \left( 2\eta_{\mu\nu} - 6w_{(\mu} n_{\nu)} + 3 n_{(\mu} n_{\nu)} \right) n^k .
\end{equation}
Using block matrix notation:
\[\eta_{\mu\nu} = \begin{pmatrix}
    1 & 0 \\ 0 & -\delta_{ij} \end{pmatrix} \ , \ 
    w_{(\mu} w_{\nu)} = \begin{pmatrix}
    1 & 0 \\ 0 & 0 \end{pmatrix}\]
\begin{equation}
     w_{(\mu} n_{\nu)} = \begin{pmatrix}
    1 & n_j/2 \\ n_i/2 & 0 \end{pmatrix} \ , \ n_{(\mu} n_{\nu)} = \begin{pmatrix}
    1 & n_j\\ n_i & n_in_j \end{pmatrix} ,
\end{equation}
we obtain:
\begin{equation} \label{G_single_rest}
   G^{0}_{\mu\nu} = \frac{M}{r_0^2} \begin{pmatrix}
    0 & n_j \\ n_i & 2\delta_{ij} -4n_i n_j \end{pmatrix} \ , \ G^{k}_{\mu\nu} =  \frac{Mn^k}{r_0^2} \begin{pmatrix}
    -1 & 0 \\ 0 & -2\delta_{ij} +3n_i n_j \end{pmatrix} .
\end{equation}

For an object in the gravitational field of the extended body, the operator $H$ is obtained by applying equation \eqref{H_operator} to the deviation tensor $h_{\mu\nu}$ of the extended body, given in equation \eqref{h_ball} (derived in Section 3). In the time-polar coordinates \eqref{coordTimePolar}, this yields:
\begin{equation}\label{Hball}
  H_\mu^\beta =
\begin{pmatrix}
\phi & \phi-\psi-\chi & 0 \\
-\phi+\psi+\chi & -\phi+2\psi & 0 \\ 
0 & 0 & -\psi
\end{pmatrix} .
\end{equation}
It remains, therefore, to determine the gravitational tensor $G^\alpha_{\mu\nu}$ for the extended body. To compute this tensor, as in Section 3, we first decompose the ball $B$ into spherical shells, and then decompose each shell into differential mass elements. The gravitational tensor $G'^\alpha_{\mu\nu}$ of a single shell is obtained by superposing (integrating) the corresponding tensors of the differential mass elements within the shell. The gravitational tensor of the entire ball, $G^\alpha_{\mu\nu}$, is then obtained by integrating the tensors of the individual shells. The gravitational tensor for a point-like mass element is given by equation \eqref{G_near}. The full calculation is provided in Appendix IV.

Due to the spherical symmetry, 
%the motion is confined to a two-dimensional plane spanned by the direction from the center of the source to the object and the object's velocity. Thus, 
we work in the time-polar coordinates defined in \eqref{coordTimePolar}. In this coordinate system, the gravitational tensor takes the form:
\[G^{0} = -\frac{M}{\rho^2} 
\begin{pmatrix}
0 & 1 & 0\\
1 & 2 & 0 \\
0 & 0 & -2
\end{pmatrix} + \frac{M_4}{\rho^4}
\begin{pmatrix}
0 & 0 & 0\\
0 & 2 & 0\\
0 & 0 & -2/3 \\
\end{pmatrix} + 
\begin{pmatrix}
0 & 0 & 0\\
0 & B(\rho) & 0\\
0 & 0 & -A(\rho)\\
\end{pmatrix}
\]
\[G^{1} = -\frac{M}{\rho^2} 
\begin{pmatrix}
1 & 0 & 0\\
0 & -1 & 0\\
0 & 0 & 2
\end{pmatrix} + \frac{M_4}{\rho^4}  \begin{pmatrix}
0 & 0 & 0
\\ 0 & -2 & 0\\
0 & 0 & 1
\end{pmatrix}\]
\begin{equation} \label{G ball 3D}
    G^{2} = \frac{M_4}{\rho^4} 
\begin{pmatrix}
0 & 0 & 0\\
0 & 0 & 1\\
0 & 1 & 0
\end{pmatrix}
\end{equation}
where
\[ A(\rho) = \sum_{n=2}^{\infty} \frac{2}{(2n-1)(2n+1)\rho^{2n+2}} M_{2n+2} ,\]
\begin{equation}
     B(\rho) = \sum_{n=2}^{\infty} \frac{2}{(2n-1)\rho^{2n+2}} M_{2n+2}.
\end{equation}
$A(\rho)$ and $B(\rho)$ contain the mass moments with order 6 and higher.
\\

The inverse matrix  $(I-H)^{-1}$ can be obtained using equation \eqref{metricBallMatrix}.
\begin{equation} \label{inverse matrix ball}
    (I - H)^{-1} = \frac{1}{D(\rho)} \begin{pmatrix}
 1+\phi -2\psi & \phi -\psi-\chi & 0 \\ -\phi +\psi + \chi & 1 - \phi &  0 \\ 0 & 0 & 0
\end{pmatrix} + \begin{pmatrix}
0 & 0 & 0 \\ 0 & 0 & 0 \\ 0 & 0 & \frac{1}{1+\psi}
\end{pmatrix}
\end{equation}
where $D(\rho)$ is the determinant of the $2\times 2$ matrix
\begin{equation} \label{D}
    D(\rho) = 1-2\psi(1-\phi)+(\chi+\psi)^2-2\phi (\chi+\psi) .
\end{equation}

In our coordinates, the 4-velocity of the object is $\dot{x}^\alpha = (1,\beta_r,\beta_\perp)$. 
The time acceleration is $ \ddot{x}^{0} =0$, as expected.

The radial and transverse accelerations are obtained by substituting $G^\alpha_{\mu\nu}$, $(I - H)^{-1}$, and $\dot{x}^\alpha$ into \eqref{acceleration_tau}. 

The tensor $G$ contains contributions of order $\frac{M}{\rho^2}$, corresponding to a point-like source, order $\frac{M_4}{\rho^4}$, representing the leading contribution from the extended body, and higher-order terms appear only in the last matrix of $G^0$.
Accordingly, the acceleration can be decomposed into contributions of different orders.

To compute the acceleration, $G^\alpha_{\mu\nu}$ must be contracted twice with the four-velocity $\dot{x}^\mu$. Since all components of the extended body contributions to $G$ involving the time index (0) vanish, the correction to the acceleration from the extended structure is quadratic in $\beta$.
As a result, these corrections are relatively small for non-relativistic velocities.

\section{Summary and Discussion}

In this paper, we employed the framework of Extended Relativity (ER) to describe the relativistic gravitational field of an extended spherically symmetric body and to identify deviations from the Newtonian shell theorem.

In the Introduction we point out the main ideas of ER and compared them to other models unifying gravity and electromagnetism.

In Section 2, we introduced the ER metric \eqref{metricSSstat} for a single point-like source. We present the deviation of the local spacetime geometry caused by the gravitational field by viewing the ball of admissible velocities near each point.  Figure \ref{BH_new} illustrates the ball of admissible velocities for objects at various distances from the event horizon. This influence is physically intuitive.

In Section 3, we describe how to construct the combined metric \eqref{devComb} for a gravitational field generated by multiple sources.
We applied this construction to derive the metric \eqref{metricBall} for an extended spherically symmetric mass distribution. Notably, we found that the gravitational time dilation produced by an extended body is identical to that of a point-like source. We observed that Newtonian shell theorem does not hold exactly, but holds approximately, especially in the far region. 

In Section 4, we defined the balls of admissible velocities \eqref{velocity_vall_extended_eq} in the fields of extended bodies. We compared them to those arising from point-like sources, focusing on the surfaces of a neutron star (Subsection \ref{neutron}) and the Earth (Subsection \ref{Earth}). In Subsection \ref{roundtrip}, we analyzed the effect of the extended mass distribution on round-trip times between the Earth's surface and the ISS or the Moon. We found that the correction due to the extended body model is of the same order of magnitude as the relativistic correction from a single point source.

In Section 5, we derived explicit expressions for the tensors used in calculating the acceleration of an object in the gravitational field of an extended, spherically symmetric body.

We conclude that, although the Newtonian shell theorem does not hold exactly within the ER framework, it remains a highly accurate approximation for most practical applications.

The analysis developed in this paper may improve the navigation of satellites in strongly elliptical orbits by incorporating relativistic corrections. The methods could be helpful to derive the relativistic gravitational field of a rotating spherically symmetric body. This may help reveal relativistic phenomena in neutron stars and pulsars.

\section{Acknowledgements}
We thank the referees for important comments that were very helpful in improving the article.
This work includes results from the MSc thesis of the second-named author, conducted under the supervision of the first-named author. The authors also acknowledge the use of ChatGPT for assistance with English language editing.

\newpage

\section{Appendix I. Deviation Tensor of the Extended Body}

In this appendix, we will calculate the deviation tensor \( h_{\mu\nu} \) for the spherically symmetric body discussed in section 5.

First we will calculate deviation tensor \( h'_{\mu\nu} \) for a spherical shell. To do so, we will evaluate the integrals in equation \eqref{H_integral_equation}, integrating first by $ d\varphi'$ and then by $d\theta'$. For majority of entries the first integration give zero. So, we need to calculate the non-zero terms. All of them are calculated  by replacing the integration by  $\theta '$ with integration by $g =\sqrt{q^2+1-2q \cos\theta'}$ where
$q=\rho'/\rho,\;\;0<q<1 \,$.  For this, by differentiating equation $g^2=q^2+1-2q \cos\theta'$ by $g$ we obtain
\(2g =2q \sin \theta' \frac{d\theta'}{dg}\) or
\begin{equation}
     \sin \theta' d\theta'   = \frac{g}{q}dg .
\end{equation}
Replacing the integration by  $\theta '$ with $g$, the limits $[ 0,\pi]$ become $[1-q,1+q])$. 

\begin{equation}
    l_\mu l_\nu = \frac {1} {\rho g(q,\theta')} \cdot \begin{pmatrix}
1 & n_i \\ n_j & n_i n_j
\end{pmatrix} = \frac{1} {\rho g^3(q,\theta')} \cdot
\end{equation}
\[ \left(
\begin{smallmatrix}
g^2 & -gq \sin \theta'\cos\varphi' & -gq \sin \theta'\sin\varphi' & g (1-q\cos \theta') \\
-g q\sin \theta'\cos\varphi' &  q^2\sin^2 \theta'\cos^2\varphi' &  q^2\sin^2 \theta'\cos\varphi' \sin\varphi' & -q\sin \theta'\cos\varphi'  (1-q\cos \theta') \\
-g q\sin \theta'\sin\varphi'  &  q^2\sin^2 \theta'\sin\varphi' \cos\varphi'  & q^2 \sin^2 \theta'\sin^2\varphi' & -q\sin \theta'\sin\varphi'  (1-q\cos \theta') \\
g (1-q\cos \theta') &- q(1-q\cos \theta') \sin \theta'\cos\varphi' & -q(1-q\cos \theta')\sin \theta'\sin\varphi' & (1-q\cos \theta')^2
\end{smallmatrix}
\right)
\]
\\

\begin{equation}
    h'_{00} = \frac{2M'}{4\pi} \int_{0}^{2\pi} \int_{0}^{\pi}  \frac { \sin\theta'd\theta'} {\rho g(q,\theta')} d\varphi'=\frac{M'}{\rho}  \int_{0}^{\pi} \frac { \sin\theta'd\theta'} { g(q,\theta')}.
\end{equation}
Replacing the integration by  $\theta '$ with $g$,yields, 
\begin{equation}
   h'_{00} = \frac{M'}{\rho} \int_{0}^{\pi} \frac { \sin\theta'd\theta'} { g(q,\theta')}  =\frac{M'}{\rho}\int_{1-q}^{1+q} \frac{1}{g} \frac{g}{q}dg = \frac{M'}{\rho} \frac{2q}{q}=\frac{2M'}{\rho}.
\end{equation}

Integrating the expressions for $h'_{11},\;h'_{22}$ by $\varphi'$ we obtain
  \begin{equation} h'_{11} =h'_{22} = \frac{M'q^2}{2\rho} \int_{0}^{\pi}  \frac {\sin^2 \theta'} {g^3 } \sin\theta'  d\theta' .
\end{equation}
Using $\sin^2 \theta' = 1-\cos^2 \theta'$ we obtain
\[\int_{0}^{\pi} \frac { \sin^2 \theta'} {g^3 } \sin\theta'  d\theta'=\int_{1-q}^{1+q} \frac{1-\left(\frac{q^2+1-g^2}{2q}\right)^2}{g^3} \frac{g}{q}dg =\]
\[\frac{-1}{4q^3} \int_{1-q}^{1+q} \frac{(q^2-1)^2-2g^2(q^2+1)+g^4}{g^2} dg = \frac{-1}{4q^3} \left( (q^2-1)^2 \left[ \frac{1}{1-q} - \frac{1}{1+q} \right] \right.\]
\[\left. - 2(q^2+1) \left[ 1+q - (1-q) \right] + \frac{(1+q)^3}{3} - \frac{(1-q)^3}{3} \right) =\]
\[\frac{-1}{4q^3} \left( 2q(1-q^2) - 4q(q^2+1) + \frac{2q}{3} (q^2+3)\right) = \frac{4}{3}\]
implying that
\begin{equation}
    h'_{11} = \frac{M'q^2}{2\rho}  \frac {4}{3}   = \frac{2M'q^2}{3\rho } .
\end{equation} 

\begin{equation}
    h'_{33} = \frac{2M'}{4\pi} \int_{0}^{2\pi} \int_{0}^{\pi}  \frac {(1-q\cos\theta')^2} {\rho g(q,\theta')^3} \sin\theta'  d\theta' d\varphi'=\frac{M'}{\rho}  \int_{0}^{\pi}  \frac {(1-q\cos\theta')^2} {g(q,\theta')^3} \sin\theta'  d\theta' .
\end{equation}
\[ \int_{0}^{\pi} \frac {(1-q\cos\theta')^2} {g^3} \sin\theta' d\theta' = \int_{1-q}^{1+q} \frac{\left(1 - \frac{q^2+1-g^2}{2}\right)^2}{g^3} \frac{g}{q}dg =\]
\[\frac{1}{4q} \int_{1-q}^{1+q} \frac{\left(q^2-1-g^2\right)^2}{g^2} dg = \frac{1}{4q} \int_{1-q}^{1+q} \frac{(q^2-1)^2-2g^2(q^2-1)+g^4}{g^2} dg = \]
\[\frac{1}{4q} \left( (q^2-1)^2 \left[ \frac{1}{1-q} - \frac{1}{1+q} \right] - 2(q^2-1) \left[ 1+q - (1-q) \right] \right.\]
\[\left. + \frac{(1+q)^3}{3} - \frac{(1-q)^3}{3} \right) = \frac{1}{4q} \left( 2q(1-q^2) - 4q(q^2-1) + \frac{2q}{3} (q^2+3)\right)\]
\[= \frac{2q}{4q} \left( 4- \frac{8q^2}{3} \right) = 2 - \frac{4q^2}{3} \]
Therefore
\begin{equation}
    h'_{33} = \frac{2M'}{\rho} \left(1 - \frac{2q^2}{3} \right).
\end{equation}

\[h'_{03} =h'_{30} =\frac{2M'}{4\pi} \int_{0}^{2\pi} \int_{0}^{\pi}  \frac {(1-q\cos\theta') \sin\theta'  d\theta' } {\rho g(q,\theta')^2}d\varphi'=\]
\begin{equation}
   =\frac{M'}{\rho}  \int_{0}^{\pi} \frac {(1-q\cos\theta') \sin\theta'  d\theta' } { g(q,\theta')^2}.
\end{equation}
Since \[ \int_{0}^{\pi}\frac {(1-q\cos\theta') \sin\theta'  d\theta' } { g^2}=\int_{1-q}^{1+q} \frac{1 - \frac{q^2+1-g^2}{2}}{g^2} \frac{g}{q}dg = \]
\begin{equation}
   \frac{1-q^2}{2q} \int_{1-q}^{1+q} \frac{dg}{g} + \frac{1}{2q} \int_{1-q}^{1+q} g dg =  \frac{1-q^2}{2q}\ln{\frac{1+q}{1-q}} + 1 .
\end{equation}

For later calculation we need to expand the last function into a power series in $q$, with $0<q<1$. 
Using the Taylor expansion on $\ln(1+x)$ 
\begin{equation}
    \ln{\frac{1+q}{1-q}}  =   
\sum_{n=0}^{\infty} \frac{2q^{2n+1}}{2n+1} . 
\end{equation}
Thus,
\[\frac{1-q^2}{2q}\ln{\frac{1+q}{1-q}}=\sum_{n=0}^{\infty} \frac{1}{2n+1} q^{2n}  - \sum_{n=0}^{\infty} \frac{1}{2n+1} q^{2n+2} =\]
\[=\sum_{n=0}^{\infty} \frac{q^{2n}}{2n+1}  - \sum_{n=1}^{\infty} \frac{q^{2n}}{2n-1} =\sum_{n=1}^{\infty} \frac{-2q^{2n}}{(2n-1)(2n+1)}+1 \]

From this we obtain
\begin{equation}
   h'_{03} =h'_{03}=\frac{2M'}{\rho}\left(1 -b(q) \right),\;\; b(q)= \sum_{n=1}^{\infty} \frac{q^{2n}}{(2n-1)(2n+1)}
\end{equation}

\begin{equation}
    h'_{03} \approx \frac{2M'}{\rho} \left( 1 - \frac{q^2}{3} + O(q^4)\right)
\end{equation}

\newpage

\section{Appendix II - MATLAB Simulation - Round Trip Time}

\begin{lstlisting}[language=Matlab]
rs = 8.87*1e-3; % Earth's schwarzschild radius
M=rs/2;
R = 6371*1e3; %Earth surface
r = R + 400e3; %ISS
c=3e8;

dx = (r-R)/1e4;
%forward - radially outward
t1_extended = 0; %Extended body
t1_single = 0; % ER single point source
t1_Schwar = 0; % Schwarzschild
for x=R:dx:(r-dx)
   phi = 2*M/x;
   psi = (2*M/x)*((R/x)^2)/5;
   chi = 0;
   for n=2:1:20
       chi = chi + (2*M/x) * 3*((R/x)^(2*n))/((2*n-1)*(2*n+1)*(2*n+3));
   end
   beta_single = (1-phi) / (1+phi);
   b_a = (phi-psi-chi)/(1+phi-2*psi);
   beta_extended = sqrt((1-phi)/(1+phi-2*psi) + (b_a)^2) - b_a;
   beta_Schwar = 1-phi; 
   t1_extended = t1_extended + dx/beta_extended;
   t1_single =t1_single + dx/beta_single;
   t1_Schwar = t1_Schwar + dx/beta_Schwar;
end


%backward - radially toward
t2_extended = 0;
t2_single = 0;
for x=(r-dx):-dx:R
   phi = 2*M/x;
   psi = (2*M/x)*((R/x)^2)/5;
   chi = 0;
   for n=2:1:20
       chi = chi + (2*M/x) * 3*((R/x)^(2*n))/((2*n-1)*(2*n+1)*(2*n+3));
   end
   beta_single = -1;
   b_a = (phi-psi-chi)/(1+phi-2*psi);
   beta_extended = -sqrt((1-phi)/(1+phi-2*psi) + (b_a)^2) - b_a;
   t2_extended = t2_extended - dx/beta_extended;
   t2_single =t2_single - dx/beta_single;
end

%%
t1_extended = t1_extended/c; t2_extended = t2_extended/c;
t1_single = t1_single/c; t2_single = t2_single/c;
t1_Schwar = t1_Schwar/c; t2_Schwar = t1_Schwar;
t1_classical = (r-R)/c; t2_classical = t1_classical;

%%

dt1_single = t1_single-t1_classical
dt2_single = t2_single-t2_classical

dt1_Schwar = t1_Schwar-t1_classical
dt2_Schwar = t2_Schwar-t2_classical

dt1_extended = t1_extended-t1_classical
dt2_extended = t2_extended-t2_classical
\end{lstlisting}

\newpage

\section{Appendix III - Transformation from Arc-Length Parameter $s$ to Coordinate Time $t$}

Starting from the geodesic equation \eqref{acceleration_s} in the form
\begin{equation} \label{geodesic_s}
\frac{d^2x^\alpha}{ds^2} = -\Gamma^\alpha_{\mu\nu} \frac{dx^\mu}{ds} \frac{dx^\nu}{ds} ,
\end{equation}
we wish to express the motion in terms of the coordinate time $t = x^0$ instead of the arc-length parameter $s$.
Applying the chain rule gives
\begin{equation}
\frac{dx^\mu}{ds} = \frac{dx^\mu}{dt} \frac{dt}{ds}, \quad
\frac{d^2x^\alpha}{ds^2} = \frac{d^2x^\alpha}{dt^2} \left(\frac{dt}{ds}\right)^2 + \frac{dx^\alpha}{dt} \frac{d^2t}{ds^2}.
\end{equation}
Substituting this into \eqref{geodesic_s} and multiplying  both sides by $\left(\frac{ds}{dt}\right)^2$, we obtain
\begin{equation} \label{geodesic_t1}
\frac{d^2x^\alpha}{dt^2} + \frac{dx^\alpha}{dt} \frac{d^2t}{ds^2} \left(\frac{ds}{dt}\right)^2 = -\Gamma^\alpha_{\mu\nu} \frac{dx^\mu}{dt} \frac{dx^\nu}{dt}.
\end{equation}
Since $t = x^0$, it follows that $\frac{dx^0}{dt} = 1$ and $\frac{d^2x^0}{dt^2} = 0$. Substituting $\alpha = 0$ in \eqref{geodesic_t1} yields:
\begin{equation}
\frac{d^2t}{ds^2} \left(\frac{ds}{dt}\right)^2 = -\Gamma^0_{\mu\nu} \frac{dx^\mu}{dt} \frac{dx^\nu}{dt}.
\end{equation}
Substituting this result back into \eqref{geodesic_t1} yields the equation of motion with respect to coordinate time:
\begin{equation} \label{geodesic_t}
\frac{d^2x^\alpha}{dt^2} = -\Gamma^\alpha_{\mu\nu} \frac{dx^\mu}{dt} \frac{dx^\nu}{dt} + \Gamma^0_{\mu\nu} \frac{dx^\mu}{dt} \frac{dx^\nu}{dt} \frac{dx^\alpha}{dt}.
\end{equation}

We now denote derivatives with respect to $t$ by a dot, i.e., $\dot{x}^\mu = \frac{dx^\mu}{dt}$, and define the \textit{quadratic four-acceleration} as
\begin{equation}
\ddot{x}^\alpha_{[2]} = \left( (I - H)^{-1} \right)^\alpha_{\ \lambda} G^\lambda_{\mu\nu} \dot{x}^\mu \dot{x}^\nu.
\end{equation}
With this notation, the equation of motion \eqref{geodesic_t} takes the compact form
\begin{equation} 
\ddot{x}^\alpha = \ddot{x}^\alpha_{[2]} - \ddot{x}^0_{[2]} \dot{x}^\alpha,
\end{equation}
which is exactly equation \eqref{acceleration_tau} in the main text.

\newpage

\section{Appendix IV - The Gravitational Field Tensor $G^{\alpha}_{\mu\nu}$ of the Extneded Body}

In this appendix, we will calculate the gravitational field tensor \( G^\alpha_{\mu\nu} \) for the spherically symmetric body discussed in section 7. First we will calculate gravitational tensor \( G'^\alpha_{\mu\nu} \) for a spherical shell.
To do so, we will integrate the gravitational field tensor for a point-like differential mass element over the shell.

For the case of figure \ref{fig:shell}
\begin{equation}
    r_\alpha = \rho(g(q,\theta'), q\sin\theta' \cos \varphi' , q\sin\theta' \sin \varphi',q\cos\theta' -1) 
\end{equation}
where
\begin{equation}
g(q, \theta') = |\mathbf{r}| = \sqrt{q^2 + 1 - 2q \cos\theta'} ,
\end{equation}
and
\begin{equation}
    n_\alpha = \frac{r_\alpha}{r_0} = \left(1 \ , \ \frac{q}{g}\sin\theta' \cos \varphi' \ ,  \ \frac{q}{g}\sin\theta' \sin \varphi' \ , \ \frac{q}{g}\cos\theta' -\frac{1}{g} \right)
\end{equation}

Since our source is at rest, $w^\alpha=(1,0,0,0)$. Thus, $r\cdot w = \rho$ .

To calculate the gravitational tensor $G^\alpha_{\mu\nu}$, we use equation \eqref{G_single_rest}
\begin{equation}
   G^{0} = \frac{M}{r_0^2} \begin{pmatrix}
    0 & n_j \\ n_i & 2\delta_{ij} -4n_in_i \end{pmatrix} \ , \ G^{k} =  \frac{Mn^k}{r_0^2} \begin{pmatrix}
    -1 & 0 \\ 0 & -2\delta_{ij} +3n_in_j \end{pmatrix} .
\end{equation}
Notice that $G^\alpha_{\mu\nu}$ is a symmetric tensor in $\mu$ and $\nu$.

 As in Appendix II, we first integrate by $ d\varphi'$ and then by $d\theta'$. The integrations by $\theta$ are calculated by replacing the integration by  $\theta '$ with integration by $g =\sqrt{q^2+1-2q \cos\theta'}$ where
$q=\rho'/\rho,\;\;0<q<1 \,$.  For this, by differentiating equation $g^2=q^2+1-2q \cos\theta'$ by $g$ we obtain
\(2g =2q \sin \theta' \frac{d\theta'}{dg}\) or
\begin{equation}
     \sin \theta' d\theta'   = \frac{g}{q}dg .
\end{equation}
Replacing the integration by  $\theta '$ with $g$, the limits $[ 0,\pi]$ become $[1-q,1+q])$.

All the components of the tensor that involve function of $\varphi$, that the integrals from $0$ to $2\pi$ is zero, such as $\cos\varphi'$ , $\cos\varphi'\sin\varphi'$ , etc. , these components vanish.
The non-zero components are

\[G'^0_{03} = \frac{M'}{4\pi} \int_{0}^{2\pi} \int_{0}^{\pi}  \frac { q\cos\theta' -1} {\rho^2 g^3(q,\theta')} \sin\theta'd\theta' d\varphi' = \frac{M'}{2\rho^2} \int_{0}^{\pi}  \frac { q\cos\theta' -1} { g^3(q,\theta')} \sin\theta'd\theta\]

\[\int_{0}^{\pi} \frac{q\cos\theta'-1}{g^3(q,\theta')} \sin\theta' d\theta'
= \int_{1-q}^{1+q} \frac{ \frac{q^2+1-g^2}{2}-1}{g^3} \frac{g}{q}dg =\]
\[\frac{1}{2q} \int_{1-q}^{1+q} \frac{q^2-1-g^2}{g^2} dg = \frac{1}{2q} \left( (q^2-1) \left[ \frac{1}{1-q} - \frac{1}{1+q} \right] - \left[ 1+q - (1-q) \right] \right) \]
\[ = \frac{1}{2q} \left( -2q -2q \right) = -2 \]
Therefore,
\begin{equation}
    G'^0_{03} = \frac{-M'}{\rho^2} . 
\end{equation}

\[G'^0_{11} = G'^0_{22} = \frac{M'}{4\pi} \int_{0}^{2\pi} \int_{0}^{\pi}  \frac { 2g^2(q,\theta') - 4q^2 \sin^2\theta' \cos^2\varphi'} {\rho^2 g^4(q,\theta')} \sin\theta'd\theta' d\varphi' = \]
\[ \frac{M'}{\rho^2} \int_{0}^{\pi}  \frac { g^2(q,\theta') - q^2 \sin^2\theta' } {g^4(q,\theta')} \sin\theta'd\theta'  \]
Using $\sin^2 \theta' = 1-\cos^2 \theta'$ we obtain
\[\int_{0}^{\pi}  \frac{ g^2(q,\theta') - q^2 \sin^2\theta' } {g^4(q,\theta')} \sin\theta'd\theta' = \int_{1-q}^{1+q} \frac{g^2 - q^2\left( 1-\left(\frac{q^2+1-g^2}{2q}\right)^2 \right) }{g^4} \frac{g}{q}dg =\]
\[\frac{1}{4q} \int_{1-q}^{1+q} \frac{(q^2-1)^2-2g^2(q^2-1)+g^4}{g^3} dg = \frac{1}{4q} \left( \frac{(q^2-1)^2}{2} \left[ \frac{1}{(1-q)^2} - \frac{1}{(1+q)^2} \right] \right.\]
\[\left. - 2(q^2-1) \ln \left( \frac{1+q}{1-q}\right) + \frac{(1+q)^2}{2} - \frac{(1-q)^2}{2} \right) =\]
\[\frac{1}{4q} \left( 2q - 2(q^2-1) \ln \left( \frac{1+q}{1-q}\right) + 2q \right) = 1 + \frac{1-q^2}{2q} \ln \left( \frac{1+q}{1-q}\right)\]

This term was already calculated for $h'_{03}$ in Appendix II. We get
\[G'^0_{11} = G'^0_{22} =\frac{2M'}{\rho^2}\left(1 -b(q) \right),\;\; b(q)= \sum_{n=1}^{\infty} \frac{q^{2n}}{(2n-1)(2n+1)}\]
\begin{equation}
    G'^{0}_{11} =G'^{0}_{22}= \frac{2M'}{\rho^2} \left(1 - \frac{q^2}{3} + O\left(q^{4}\right) \right)
\end{equation}

\[G'^0_{33} = \frac{M'}{4\pi} \int_{0}^{2\pi} \int_{0}^{\pi}  \frac { 2g^2(q,\theta') - 4(q\cos\theta' - 1)^2} {\rho^2 g^4(q,\theta')} \sin\theta'd\theta' d\varphi' = \]
\[ \frac{M'}{\rho^2} \int_{0}^{\pi}  \frac { g^2(q,\theta') - 2(q\cos\theta' - 1)^2 } {g^4(q,\theta')} \sin\theta'd\theta'  \]

\[\int_{0}^{\pi}  \frac { g^2(q,\theta') - 2(q\cos\theta' - 1)^2 } {g^4(q,\theta')} \sin\theta'd\theta'
= \int_{1-q}^{1+q} \frac{ g^2 -2\left(\frac{q^2+1-g^2}{2}-1 \right)^2}{g^4} \frac{g}{q}dg =\]
\[\frac{1}{2q} \int_{1-q}^{1+q} \frac{-(q^2-1)^2+2g^2q^2-g^4}{g^3} dg = \frac{1}{2q} \left( \frac{-(q^2-1)^2}{2} \left[ \frac{1}{(1-q)^2} - \frac{1}{(1+q)^2} \right] \right. \]
\[ \left. + 2q^2\ln \left( \frac{1+q}{1-q}\right) - \frac{(1+q)^2}{2} + \frac{(1-q)^2}{2} \right) = \]
\[\frac{1}{2q} \left( -2q + 2q^2 \ln \left( \frac{1+q}{1-q}\right) - 2q \right) = -2 + q \ln \left( \frac{1+q}{1-q}\right)\]

As before, we expand the last function into a power series in $q$, with $0<q<1$. 
Using the Taylor expansion on $\ln(1+x)$ 
\begin{equation}
    \ln{\frac{1+q}{1-q}}  =   
\sum_{n=0}^{\infty} \frac{2q^{2n+1}}{2n+1} = \sum_{n=1}^{\infty} \frac{2q^{2n-1}}{2n-1} . 
\end{equation}
Thus,
\[q \ln \left( \frac{1+q}{1-q}\right) = 2 \sum_{n=1}^{\infty} \frac{q^{2n}}{(2n-1)} .\]
From this we obtain
\[G'^0_{33} = \frac{2M'}{\rho^2}\left(-1 + d(q) \right),\;\; d(q)= \sum_{n=1}^{\infty} \frac{q^{2n}}{(2n-1)}\]
\begin{equation}
    G'^0_{33} \approx \frac{2M'}{\rho^2} \left( -1 + q^2 + O(q^4)\right)
\end{equation}
\\

\[G'^1_{13} = \frac{-M'}{4\pi} \int_{0}^{2\pi} \int_{0}^{\pi}  \frac { 3q^2\sin^2\theta' \cos^2\varphi' (q\cos\theta' -1) } {\rho^2 g^5(q,\theta')} \sin\theta'd\theta' d\varphi' \]
\[ = \frac{-3M'}{4\rho^2} \int_{0}^{\pi}  \frac { q^2\sin^2\theta'(q\cos\theta' -1)} { g^5(q,\theta')} \sin\theta'd\theta\]
Using $\sin^2 \theta' = 1-\cos^2 \theta'$ we obtain
\[ \int_{0}^{\pi}  \frac { q^2\sin^2\theta'(q\cos\theta' -1)} { g^5(q,\theta')} \sin\theta'd\theta = \int_{1-q}^{1+q} \frac{q^2\left( 1-\left(\frac{q^2+1-g^2}{2q}\right)^2 \right) \left( \frac{q^2+1-g^2}{2} - 1 \right) }{g^5} \frac{g}{q}dg \]
\[= \frac{1}{8q} \int_{1-q}^{1+q} \frac{ \left( -(q^2-1)^2+2g^2(q^2+1)-g^4 \right) \left( q^2-1-g^2 \right) }{g^4}dg\]
\[= \frac{1}{8q} \int_{1-q}^{1+q} \frac{ -(q^2-1)^3 + g^2 \left[ (q^2-1)^2 +2(q^2+1)(q^2-1) \right] }{g^4}dg\]
\[+\frac{1}{8q} \int_{1-q}^{1+q} \frac{ - g^4 \left[ (q^2-1) +2(q^2+1) \right] +g^6 }{g^4}dg\]
\[= \frac{1}{8q} \int_{1-q}^{1+q} \frac{ -(q^2-1)^3 + g^2 (q^2-1)(3q^2+1) - g^4 (3q^2+1) +g^6}{g^4}dg\]
\[= \frac{1}{8q} \left( \frac{-(q^2-1)^3}{3} \left[ \frac{1}{(1-q)^3} - \frac{1}{(1+q)^3} \right] + (q^2-1)(3q^2+1) \left[ \frac{1}{1-q} - \frac{1}{1+q} \right] \right.\]
\[\left.  - (3q^2+1) \left( (1+q)-(1-q) \right) + \frac{(1+q)^3}{3} - \frac{(1-q)^3}{3} \right) =\]
\[\frac{1}{8q} \left( \frac{2q}{3}(3+q^2) -2q(3q^2+1) - 2q(3q^2+1) + \frac{2q}{3}(3+q^2) \right)\]
\[= \frac{4q}{8q} \left( \frac{3+q^2}{3} - (3q^2+1) \right) = \frac{-4q^2}{3}\]
From this we obtain
\begin{equation}
    G'^1_{13} = \frac{M'q^2}{\rho^2} .
\end{equation}
Similarly for $\alpha=2$
\begin{equation}
    G'^2_{23} = \frac{M'q^2}{\rho^2}
\end{equation}
\\

\[G'^3_{00} = \frac{M'}{4\pi} \int_{0}^{2\pi} \int_{0}^{\pi}  \frac { q\cos\theta' -1 } {\rho^2 g^3(q,\theta')} \sin\theta'd\theta' d\varphi' \]

This expression is identical to that of $G'^0_{03}$. Therefore, 
\begin{equation}
    G'^3_{00} = G'^3_{03} = \frac{-M'}{\rho^2} . 
\end{equation}

\[G'^3_{11} = G'^3_{22} = \frac{M'}{4\pi} \int_{0}^{2\pi} \int_{0}^{\pi}  \frac {-2( 1-q\cos\theta' )} {\rho^2 g^3(q,\theta')} \sin\theta'd\theta' d\varphi'+\]
\[\frac{M'}{4\pi} \int_{0}^{2\pi} \int_{0}^{\pi}  \frac {3q^2 \sin^2\theta' \cos^2\varphi' (1 - q\cos\theta' )} {\rho^2 g^5(q,\theta')} \sin\theta'd\theta' d\varphi' .\]
The first integral is similar to $G'^3_{00}$, and it's equal to $\frac{-2M'}{\rho^2}$.
The second integral is similar to $G'^1_{13}$, and it's equal to $\frac{M'q^2}{\rho^2}$. Therefore
\begin{equation}
    G'^3_{11} = G'^3_{22} = \frac{M'}{\rho^2} (-2+q^2) .
\end{equation}

\[G'^3_{33} = \frac{M'}{4\pi} \int_{0}^{2\pi} \int_{0}^{\pi}  \frac {-2( 1-q\cos\theta' )} {\rho^2 g^3(q,\theta')} \sin\theta'd\theta' d\varphi'+\]
\[\frac{M'}{4\pi} \int_{0}^{2\pi} \int_{0}^{\pi}  \frac {3(1 - q\cos\theta')^3} {\rho^2 g^5(q,\theta')} \sin\theta'd\theta' d\varphi' .\]
As before, the first integral is equal to $\frac{-2M'}{\rho^2}$. The second integral is
\[\frac{M'}{4\pi} \int_{0}^{2\pi} \int_{0}^{\pi}  \frac {3(1 - q\cos\theta')^3} {\rho^2 g^5(q,\theta')} \sin\theta'd\theta' d\varphi' = \frac{3M'}{2\rho^2} \int_{0}^{\pi} \frac {(1 - q\cos\theta')^3} {g^5(q,\theta')} \sin\theta'd\theta'\]
\[\int_{0}^{\pi} \frac{(1-q\cos\theta')^3}{g^5(q,\theta')} \sin\theta' d\theta'
= \int_{1-q}^{1+q} \frac{ \left(1-\frac{q^2+1-g^2}{2}\right)^3}{g^5} \frac{g}{q}dg = \]
\[\frac{1}{8q} \int_{1-q}^{1+q} \frac{(1-q^2+g^2)^3}{g^4} dg =\]
\[ \frac{1}{8q} \int_{1-q}^{1+q} \frac{(1-q^2)^3 +3g^2(1-q^2)^2+3g^4(1-q^2)+g^6}{g^4} dg\]
\[\frac{1}{8q} \left( \frac{(1-q^2)^3}{3} \left[ \frac{1}{(1-q)^3} - \frac{1}{(1+q)^3} \right] + 3(1-q^2)^2 \left[ \frac{1}{1-q} - \frac{1}{1+q} \right] \right. \]
\[ \left.+ 3(1-q^2)\left[ 1+q - (1-q) \right] + \frac{(1+q)^3}{3} - \frac{(1-q)^3}{3} \right) = \]
\[ = \frac{1}{8q} \left( \frac{2q(3+q^2)}{3} +6q(1-q^2) + 6q(1-q^2) +\frac{2q(3+q^2)}{3} \right) = \]
\[\frac{1}{2} \left( \frac{(3+q^2)}{3} +3(1-q^2) \right) =  2 \left( 1 - \frac{2q^2}{3} \right) .\]
Therefore,
\[ G'^3_{33} = \frac{-2M'}{\rho^2} + \frac{3M'}{\rho^2}  \left( 1 - \frac{2q^2}{3} \right) \]
\begin{equation}
     G'^3_{33} = \frac{M'}{\rho^2}  \left( 1 - 2q^2 \right)
\end{equation}
\\

Hence, we obtain

\[G'^{0}_{\mu\nu} = -\frac{M'}{\rho^2}
\begin{pmatrix}
0 & 0 & 0 & 1
\\ 0 & 2(-1+b(q)) & 0 & 0 \\
0 & 0 & 2(-1+b(q)) & 0 \\
1 & 0 & 0 & 2(1-d(q))
\end{pmatrix}
\]
\[G'^{1}_{\mu\nu} = -\frac{M'}{\rho^2} 
\begin{pmatrix}
0 & 0 & 0 & 0
\\ 0 & 0 & 0 & -q^2 \\
0 & 0 & 0 & 0 \\
0 & -q^2 & 0 & 0
\end{pmatrix}
\]
\[G'^{2}_{\mu\nu} = -\frac{M'}{\rho^2}
\begin{pmatrix}
0 & 0 & 0 & 0
\\ 0 & 0 & 0 & 0 \\ 0 & 0 & 0 & -q^2 \\
0 & 0 & -q^2  & 0
\end{pmatrix}
\]
\[G'^{3}_{\mu\nu} = -\frac{M'}{\rho^2}
\begin{pmatrix}
1 & 0 & 0 & 0
\\ 0 & 2 - q^2 & 0 & 0 \\
0 & 0 & 2 - q^2 & 0 \\
0 & 0 & 0 & -1 + 2q^2
\end{pmatrix}
\]

where $ b(q)= \sum_{n=1}^{\infty} \frac{q^{2n}}{(2n-1)(2n+1)}$ and $ d(q)= \sum_{n=1}^{\infty} \frac{q^{2n}}{(2n-1)}$ .
\\

To obtain the gravitational tensor $G^\alpha_{\mu\nu}$ for the entire ball $B$ of radius $R$ and mass density $\sigma_m(\rho')$, we divide the ball into spherical shells, and integrate the corresponding tensor for a shell all over the shells. Using that The differential mass of a shell between radius $\rho'$ and $\rho'+d\rho'$ is $dM'(\rho ')=\sigma_m(\rho')  4\pi\rho'^2 d\rho'$ , and the definition of the mass moment $M_n$ \eqref{Mn}, we get,
\[\int_0^R dM'(\rho')q^k=\int_0^R \sigma_m(\rho')  4\pi\rho'^2\frac{(\rho')^k}{\rho^k} d\rho'=\frac{M_{k+2}}{\rho^k} .\]

The gravitational field tensor for the ball $B$ is

\[G^{0}_{\mu\nu} = -\frac{M}{\rho^2}
\begin{pmatrix}
0 & 0 & 0 & 1
\\ 0 & -2 & 0 & 0 \\
0 & 0 & -2 & 0 \\
1 & 0 & 0 & 2
\end{pmatrix} + \frac{M_4}{\rho^4}
\begin{pmatrix}
0 & 0 & 0 & 0
\\ 0 & -2/3 & 0 & 0 \\
0 & 0 & -2/3 & 0 \\
0 & 0 & 0 & 2
\end{pmatrix} \]
\[ +
\begin{pmatrix}
0 & 0 & 0 & 0
\\ 0 & -A(\rho) & 0 & 0 \\
0 & 0 & -A(\rho) & 0 \\
0 & 0 & 0 & B(\rho)
\end{pmatrix}
\]
\[G'^{1}_{\mu\nu} = \frac{M_4}{\rho^4} 
\begin{pmatrix}
0 & 0 & 0 & 0
\\ 0 & 0 & 0 & 1 \\
0 & 0 & 0 & 0 \\
0 & 1 & 0 & 0
\end{pmatrix}
\ , \ G'^{2}_{\mu\nu} = \frac{M_4}{\rho^4}
\begin{pmatrix}
0 & 0 & 0 & 0
\\ 0 & 0 & 0 & 0 \\ 0 & 0 & 0 & 1 \\
0 & 0 & 1  & 0
\end{pmatrix}
\]
\[G'^{3}_{\mu\nu} = -\frac{M}{\rho^2}
\begin{pmatrix}
1 & 0 & 0 & 0
\\ 0 & 2& 0 & 0 \\
0 & 0 & 2 & 0 \\
0 & 0 & 0 & -1
\end{pmatrix} + \frac{M_4}{\rho^4}
\begin{pmatrix}
0 & 0 & 0 & 0
\\ 0 & 1 & 0 & 0 \\
0 & 0 & 1 & 0 \\
0 & 0 & 0 & -2
\end{pmatrix}
\]

where
\[ A(\rho) = \sum_{n=2}^{\infty} \frac{2}{(2n-1)(2n+1)\rho^{2n+2}} M_{2n+2} ,\]
\begin{equation}
     B(\rho) = \sum_{n=2}^{\infty} \frac{2}{(2n-1)\rho^{2n+2}} M_{2n+2}.
\end{equation}

\end{document}